\documentclass[%
 reprint,
 amsmath,amssymb,
 aps,
prl,
superscriptaddress,
]{revtex4-2}

\usepackage{graphicx}
\usepackage{dcolumn}
\usepackage{bm}
\usepackage{booktabs}
\usepackage{gensymb}
\usepackage[colorlinks=true,allcolors=blue]{hyperref}

\usepackage[caption=false]{subfig}

\usepackage{lineno}

\usepackage{color}
\begin{document}

\title{An eV-scale sterile neutrino search using eight years of atmospheric muon neutrino data from the IceCube Neutrino Observatory}




\date{\today}

\begin{abstract}
The results of a 3+1 sterile neutrino search using eight years of data from the IceCube Neutrino Observatory are presented.
A total of 305,735 muon neutrino events are analyzed in reconstructed energy-zenith space to test for signatures of a matter-enhanced oscillation that would occur given a sterile neutrino state with a mass-squared differences between 0.01\,eV$^2$ and 100\,eV$^2$.
The best-fit point is found to be at $\sin^2(2\theta_{24})=0.10$ and $\Delta m_{41}^2 = 4.5{\rm eV}^2$, which is consistent with the no sterile neutrino hypothesis with a p-value of 8.0\%.
\end{abstract}

\affiliation{III. Physikalisches Institut, RWTH Aachen University, D-52056 Aachen, Germany}
\affiliation{Department of Physics, University of Adelaide, Adelaide, 5005, Australia}
\affiliation{Dept. of Physics and Astronomy, University of Alaska Anchorage, 3211 Providence Dr., Anchorage, AK 99508, USA}
\affiliation{Dept. of Physics, University of Texas at Arlington, 502 Yates St., Science Hall Rm 108, Box 19059, Arlington, TX 76019, USA}
\affiliation{CTSPS, Clark-Atlanta University, Atlanta, GA 30314, USA}
\affiliation{School of Physics and Center for Relativistic Astrophysics, Georgia Institute of Technology, Atlanta, GA 30332, USA}
\affiliation{Dept. of Physics, Southern University, Baton Rouge, LA 70813, USA}
\affiliation{Dept. of Physics, University of California, Berkeley, CA 94720, USA}
\affiliation{Lawrence Berkeley National Laboratory, Berkeley, CA 94720, USA}
\affiliation{Institut f{\"u}r Physik, Humboldt-Universit{\"a}t zu Berlin, D-12489 Berlin, Germany}
\affiliation{Fakult{\"a}t f{\"u}r Physik {\&} Astronomie, Ruhr-Universit{\"a}t Bochum, D-44780 Bochum, Germany}
\affiliation{Universit{\'e} Libre de Bruxelles, Science Faculty CP230, B-1050 Brussels, Belgium}
\affiliation{Vrije Universiteit Brussel (VUB), Dienst ELEM, B-1050 Brussels, Belgium}
\affiliation{Dept. of Physics, Massachusetts Institute of Technology, Cambridge, MA 02139, USA}
\affiliation{Dept. of Physics and Institute for Global Prominent Research, Chiba University, Chiba 263-8522, Japan}
\affiliation{Department of Physics, Loyola University Chicago, Chicago, IL 60660, USA}
\affiliation{Dept. of Physics and Astronomy, University of Canterbury, Private Bag 4800, Christchurch, New Zealand}
\affiliation{Dept. of Physics, University of Maryland, College Park, MD 20742, USA}
\affiliation{Dept. of Astronomy, Ohio State University, Columbus, OH 43210, USA}
\affiliation{Dept. of Physics and Center for Cosmology and Astro-Particle Physics, Ohio State University, Columbus, OH 43210, USA}
\affiliation{Niels Bohr Institute, University of Copenhagen, DK-2100 Copenhagen, Denmark}
\affiliation{Dept. of Physics, TU Dortmund University, D-44221 Dortmund, Germany}
\affiliation{Dept. of Physics and Astronomy, Michigan State University, East Lansing, MI 48824, USA}
\affiliation{Dept. of Physics, University of Alberta, Edmonton, Alberta, Canada T6G 2E1}
\affiliation{Erlangen Centre for Astroparticle Physics, Friedrich-Alexander-Universit{\"a}t Erlangen-N{\"u}rnberg, D-91058 Erlangen, Germany}
\affiliation{Physik-department, Technische Universit{\"a}t M{\"u}nchen, D-85748 Garching, Germany}
\affiliation{D{\'e}partement de physique nucl{\'e}aire et corpusculaire, Universit{\'e} de Gen{\`e}ve, CH-1211 Gen{\`e}ve, Switzerland}
\affiliation{Dept. of Physics and Astronomy, University of Gent, B-9000 Gent, Belgium}
\affiliation{Dept. of Physics and Astronomy, University of California, Irvine, CA 92697, USA}
\affiliation{Karlsruhe Institute of Technology, Institut f{\"u}r Kernphysik, D-76021 Karlsruhe, Germany}
\affiliation{Dept. of Physics and Astronomy, University of Kansas, Lawrence, KS 66045, USA}
\affiliation{SNOLAB, 1039 Regional Road 24, Creighton Mine 9, Lively, ON, Canada P3Y 1N2}
\affiliation{Department of Physics and Astronomy, UCLA, Los Angeles, CA 90095, USA}
\affiliation{Department of Physics, Mercer University, Macon, GA 31207-0001, USA}
\affiliation{Dept. of Astronomy, University of Wisconsin, Madison, WI 53706, USA}
\affiliation{Dept. of Physics and Wisconsin IceCube Particle Astrophysics Center, University of Wisconsin, Madison, WI 53706, USA}
\affiliation{Institute of Physics, University of Mainz, Staudinger Weg 7, D-55099 Mainz, Germany}
\affiliation{Department of Physics, Marquette University, Milwaukee, WI, 53201, USA}
\affiliation{Institut f{\"u}r Kernphysik, Westf{\"a}lische Wilhelms-Universit{\"a}t M{\"u}nster, D-48149 M{\"u}nster, Germany}
\affiliation{Bartol Research Institute and Dept. of Physics and Astronomy, University of Delaware, Newark, DE 19716, USA}
\affiliation{Dept. of Physics, Yale University, New Haven, CT 06520, USA}
\affiliation{Dept. of Physics, University of Oxford, Parks Road, Oxford OX1 3PU, UK}
\affiliation{Dept. of Physics, Drexel University, 3141 Chestnut Street, Philadelphia, PA 19104, USA}
\affiliation{Physics Department, South Dakota School of Mines and Technology, Rapid City, SD 57701, USA}
\affiliation{Dept. of Physics, University of Wisconsin, River Falls, WI 54022, USA}
\affiliation{Dept. of Physics and Astronomy, University of Rochester, Rochester, NY 14627, USA}
\affiliation{Oskar Klein Centre and Dept. of Physics, Stockholm University, SE-10691 Stockholm, Sweden}
\affiliation{Dept. of Physics and Astronomy, Stony Brook University, Stony Brook, NY 11794-3800, USA}
\affiliation{Dept. of Physics, Sungkyunkwan University, Suwon 16419, Korea}
\affiliation{Institute of Basic Science, Sungkyunkwan University, Suwon 16419, Korea}
\affiliation{Dept. of Physics and Astronomy, University of Alabama, Tuscaloosa, AL 35487, USA}
\affiliation{Dept. of Astronomy and Astrophysics, Pennsylvania State University, University Park, PA 16802, USA}
\affiliation{Dept. of Physics, Pennsylvania State University, University Park, PA 16802, USA}
\affiliation{Dept. of Physics and Astronomy, Uppsala University, Box 516, S-75120 Uppsala, Sweden}
\affiliation{Dept. of Physics, University of Wuppertal, D-42119 Wuppertal, Germany}
\affiliation{DESY, D-15738 Zeuthen, Germany}

\author{M. G. Aartsen}
\affiliation{Dept. of Physics and Astronomy, University of Canterbury, Private Bag 4800, Christchurch, New Zealand}
\author{R. Abbasi}
\affiliation{Department of Physics, Loyola University Chicago, Chicago, IL 60660, USA}
\author{M. Ackermann}
\affiliation{DESY, D-15738 Zeuthen, Germany}
\author{J. Adams}
\affiliation{Dept. of Physics and Astronomy, University of Canterbury, Private Bag 4800, Christchurch, New Zealand}
\author{J. A. Aguilar}
\affiliation{Universit{\'e} Libre de Bruxelles, Science Faculty CP230, B-1050 Brussels, Belgium}
\author{M. Ahlers}
\affiliation{Niels Bohr Institute, University of Copenhagen, DK-2100 Copenhagen, Denmark}
\author{M. Ahrens}
\affiliation{Oskar Klein Centre and Dept. of Physics, Stockholm University, SE-10691 Stockholm, Sweden}
\author{C. Alispach}
\affiliation{D{\'e}partement de physique nucl{\'e}aire et corpusculaire, Universit{\'e} de Gen{\`e}ve, CH-1211 Gen{\`e}ve, Switzerland}
\author{N. M. Amin}
\affiliation{Bartol Research Institute and Dept. of Physics and Astronomy, University of Delaware, Newark, DE 19716, USA}
\author{K. Andeen}
\affiliation{Department of Physics, Marquette University, Milwaukee, WI, 53201, USA}
% ...

\collaboration{IceCube Collaboration}
\noaffiliation

\maketitle

\email{analysis@icecube.wisc.edu}

\section{Introduction~\label{sec:intro}}
The three-flavor massive neutrino oscillation formalism has been well-established experimentally~\cite{Tanabashi:2018oca,Esteban:2018azc,deSalas:2017kay,Capozzi:2016rtj}.
The standard paradigm has also been challenged, by several experiments exhibiting anomalous $\nu_e$ ($\bar{\nu}_e$) appearance in $\nu_\mu$ ($\bar{\nu}_\mu$) beams~\cite{Athanassopoulos:1997pv,Aguilar-Arevalo:2018gpe}.
These anomalies can be interpreted as evidence for subleading oscillations of $\nu_\mu\rightarrow\nu_e$ or $\bar{\nu}_\mu\rightarrow\bar{\nu}_e$ caused by additional neutrinos with large mass-squared differences in the range of $\Delta m^2\sim 0.1-10~{\rm eV}^2$~\cite{Abazajian:2012ys}.
On the other hand, measurements of the $Z$-boson decay width to invisible final states demonstrate that only three light neutrinos participate in weak interactions~\cite{ALEPH:2005ab}, so any additional neutrino flavor states must be non-weakly-interacting, or ``sterile.''
The simplest such model is referred to as a ``3+1'' model, where in addition to the three known mass states, a fourth heavier one is added.

\begin{figure}[b]
\centering
\includegraphics[width=0.93\columnwidth]{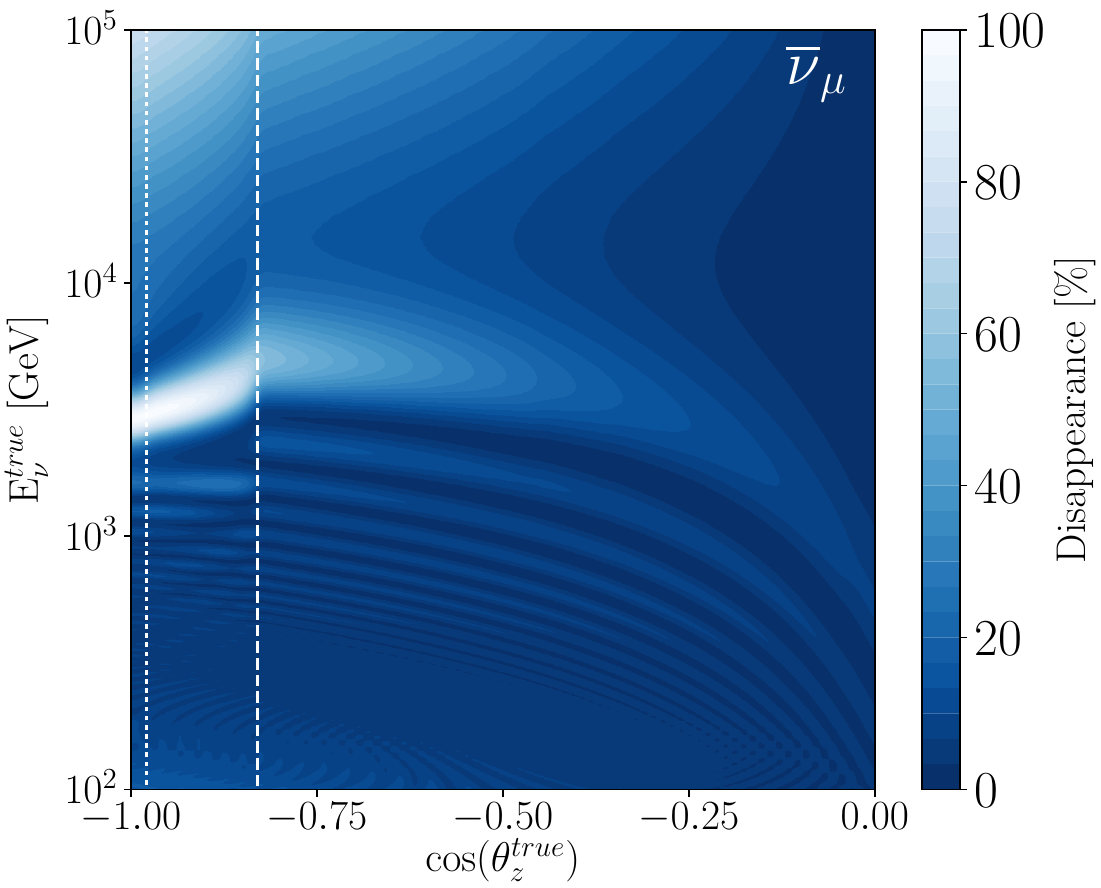}
\caption{%
\textbf{\textit{Muon-antineutrino oscillogram.}} Atmospheric $\bar\nu_\mu$ disappearance probability {\em vs.} true energy and cosine zenith at the globally preferred sterile neutrino hypothesis of Ref.~\cite{Diaz:2019fwt} ($\Delta m^2_{41}=1.3~{\rm eV}^2$, $\sin^2(2\theta_{24})=0.07$, $\sin^2(2\theta_{34})=0.0$).
Effects include a matter-enhanced resonance at TeV energies, neutrino absorption at high energy and small zenith, and vacuumlike oscillation at low energies.
The matter-enhanced resonance appears only in the antineutrino flux for the case of small angles and $\Delta m^2_{41} >0$.
Vertical white lines indicate transitions between inner to outer core ($\cos(\theta_\nu^{\rm true}) = -0.98$) and outer core to mantle ($\cos(\theta_z^{\rm true}) = -0.83$).}
\label{fig::true_signal}
\end{figure}

The relationship between the flavor and mass states is described by a unitary matrix, ${\rm U}_{PNMS}$, which in the three-neutrino model can be parameterized in terms of three mixing angles and one oscillation-accessible $CP$-violating phase.
Adding a sterile state expands the mixing matrix to four dimensions, in which the added degrees of freedom can be parameterized by introducing three new rotations with angles $\theta_{14}$, $\theta_{24}$, and $\theta_{34}$, and two new oscillation-accessible $CP$-violating phases, $\delta_{14}$ and $\delta_{24}$.
The oscillation phenomenology of the 3+1 model adds both shorter baseline vacuumlike oscillations, and also novel oscillation effects in the presence of matter~\cite{Akhmedov:1988kd,Krastev:1989ix,Chizhov:1998ug,Chizhov:1999az,Akhmedov:1999va}.
For eV-scale sterile neutrino states, for example, a matter-enhanced resonance~\cite{Nunokawa:2003ep,Choubey:2007ji,Barger:2011rc,Esmaili:2012nz,Esmaili:2013vza,Lindner:2015iaa} would result in the near complete disappearance of TeV-scale muon antineutrinos passing through the Earth's core, as shown in Fig.~\ref{fig::true_signal}.
By measuring and characterizing the flux of atmospheric neutrinos in the GeV to PeV energy range, the IceCube Neutrino Observatory is uniquely positioned to search for such matter-enhanced oscillations, a smoking-gun signature of eV-scale sterile neutrinos.

Testing the 3+1 model as an explanation of short-baseline anomalies and constraining its free parameters requires measurements in multiple oscillation channels, including $\nu_\mu\rightarrow\nu_\mu$~\cite{Dydak:1983zq,Stockdale:1984cg,Mahn:2011ea,Abe:2014gda,MINOS:2016viw,TheIceCube:2016oqi,Aartsen:2017bap,Adamson:2017uda,Albert:2018mnz}, $\nu_e\rightarrow\nu_e$~\cite{Declais:1994su,Abdurashitov:2009tn,Kaether:2010ag,Conrad:2011ce,Ko:2016owz,An:2016luf,Alekseev:2018efk,Ashenfelter:2018iov,Almazan:2018wln}, and $\nu_\mu\rightarrow\nu_e$~\cite{Athanassopoulos:1997pv,Armbruster:2002mp,Aguilar-Arevalo:2018gpe,Adamson:2008qj,Astier:2003gs}.
Fits to global data~\cite{Gariazzo:2017fdh,Dentler:2018sju,Diaz:2019fwt} find a strong preference for models with sterile neutrinos over the standard three-neutrino paradigm.
However, even at the most preferred values of $\Delta m^2 \sim  1~{\rm eV}^2$, the mixing angles required to viably explain anomalies in the $\nu_\mu\rightarrow\nu_e$ and $\bar{\nu}_\mu\rightarrow\bar{\nu}_e$ channels are in strong tension with measurements of $\nu_\mu$ and $\bar{\nu}_\mu$ disappearance. Evidence for oscillation effects beyond the three-neutrino paradigm in  $\bar{\nu}_\mu$ disappearance are yet to be observed~\cite{Dentler:2018sju}.
One of these nonobservations was made by IceCube, using a sample of 20,145 atmospheric $\nu_\mu$ and  $\bar{\nu}_\mu$ events collected over one year of detector livetime~\cite{TheIceCube:2016oqi}.

This paper updates IceCube's high-energy sterile neutrino search using an eight-year dataset and improved event selection.
The sample includes 305,735 well-reconstructed charged-current $\nu_\mu$  and $\bar\nu_\mu$  events collected from May 13$^{\mathrm{th}}$ 2011 to May 19$^{\mathrm{th}}$ 2019.
Events are binned uniformly in $\log(E^\mu_{\rm reco})$ spanning $E^\mu_\mathrm{reco} \in [500\,\mathrm{GeV}, 9976\,\mathrm{GeV}]$ in 13 bins and uniformly in $\cos \theta^\mathrm{reco}_z$ spanning the up-going region from $-1.0$ to 0.0 in 20 bins.
The event counts in each bin are used as inputs to a likelihood-based analysis to test for evidence of eV-scale sterile neutrinos.

The increased sample size of this analysis with respect to Ref~\cite{TheIceCube:2016oqi} has been accompanied by a commensurate improvement in the precision of treatments of systematic uncertainties and statistical methods.
This paper summarizes these advances and presents the main results of this search.
A companion paper, Ref.~\cite{MEOWSPRD}, contains a more detailed exposition of the technical aspects of the analysis, as well as alternate interpretations of the data in a wider space of sterile neutrino parameters.

\section{IceCube Up-going Track Sample}

\begin{figure}[t]  
    \centering
    \subfloat{\includegraphics[width=0.99\linewidth]{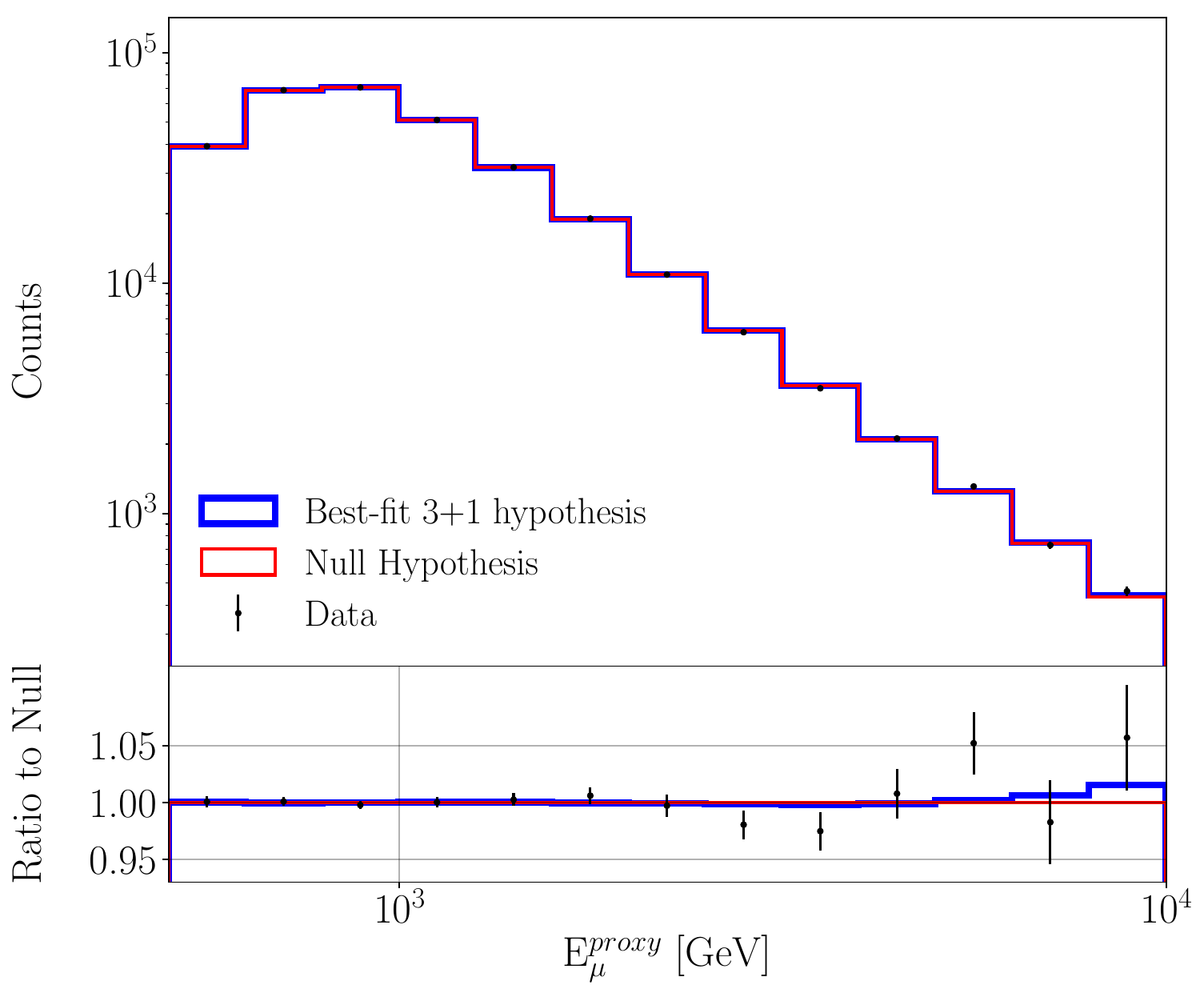}}
    \caption{\textbf{\textit{Reconstructed muon energy.}}
    Data points are shown as black markers with error bars that represent the statistical error.
    The solid blue and red lines show the best-fit sterile neutrino hypothesis and the null (no sterile neutrino) hypothesis, respectively, with nuisance parameters set to their best-fit values in each case.}
    \label{fig::event_distributions}
\end{figure} 

The IceCube Neutrino Observatory is a cubic-kilometer neutrino detector buried in the Antarctic glacier~\cite{Aartsen:2016nxy}.
It is comprised of photomultiplier tubes enclosed in glass pressure housings called ``Digital Optical Modules'' (DOMs)~\cite{Abbasi:2008aa}.
These are arranged in vertical strings on a hexagonal lattice.
The main array consists of 78 strings spaced 125\,m apart, each supporting 60 downward-facing DOMs with a 17\,m vertical spacing.
A denser array called  DeepCore~\cite{Collaboration:2011ym} instruments the clearest part of the ice within the main array.
The eight strings of DeepCore are arranged with lateral spacing between 42\,m and 72\,m and vertical DOM separation of 7\,m.
This analysis uses the complete set of IceCube DOMs in both the main array and DeepCore. 

The majority of IceCube events are produced by high-energy muons and neutrinos from cosmic-ray air showers.
Down-going ($\cos\theta_z^{\rm true}>0$) atmospheric muons (and antimuons) can penetrate the 1450\,m vertical overburden of the detector, triggering at a rate of $\sim$3\,kHz~\cite{Bugaev:1998bi}.
These events dominate the Southern-hemisphere through-going sample.
Up-going atmospheric muons, on the other hand, are effectively removed by the large overburden provided by the Earth.
Thus, muons originating from the Northern hemisphere are dominated by those produced in charged-current neutrino interactions.
A charged-current $\nu_\mu$ interaction will produce a forward secondary muon, with an energy typically between 50\% and 80\% of that of the parent $\nu_\mu$~\cite{Gandhi:1995tf}. The muon travels through the ice emitting Cherenkov radiation. 
While photons travel tens to hundreds of meters before being absorbed by the impurities in the ice~\cite{GRL:GRL20535,Askebjer:1997ep,Aartsen:2013rt}, muons with TeV energies are able to penetrate multiple kilometers of ice before falling below the Cherenkov threshold~\cite{Lipari:1993hd,Koehne:2013gpa}.
This produces an extended tracklike signature.
These events originate either inside of the detector or from a target volume  extending meters to kilometers outside the array, depending on energy~\cite{Lipari:1993hd,Halzen:2006mq}.

Events used in this analysis first pass a filter that selects muon-like events for satellite transmission to the North, and are then subject to further data-reduction techniques to reject low-energy and poorly reconstructed tracks.
Only data periods with 86 active IceCube strings and greater than 5,000 active DOMs in the detector are considered. 
A high-level event selection is applied, leveraging morphology, measures of track reconstruction quality, and the expected transmission of signal events through the zenith-dependent overburden, explained in detail in Ref.~\cite{MEOWSPRD} and based on Ref.~\cite{Aartsen:2015rwa}.
The reconstructed energy and direction of each event is calculated according to the time and geometry of light detected throughout the array~\cite{Ahrens:2003fg,Aartsen:2013vja}. The angular resolution $\sigma_{\cos \theta_z}$ varies between 0.005 and 0.015 and energy resolution of $\sigma_{\log_{10}  E_\mu}\sim 0.5$, as in the previous version of this analysis~\cite{TheIceCube:2016oqi}. 
The energy distribution of selected events is shown in Fig.~\ref{fig::event_distributions}.

Cosmic-ray muon background contamination is assessed using \texttt{CORSIKA}~\cite{Heck:1998vt},
with primary cosmic-ray energies ranging from 600\,GeV to $10^{11}$\,GeV.
Approximately 10\% of the dataset of neutrino events are predicted to contain a coincident cosmic-ray muon within the readout frame.
The $\nu_\mu$ and cosmic-ray muon tracks are separated into sub-events using an event splitter, and each sub-event is treated independently in the event selection.
After splitting and event selection, the sample is predicted to be $>99.9\%$ pure in $\nu_\mu / \bar{\nu}_\mu$ induced events~\cite{MEOWSPRD}.

\section{Sterile Neutrino Analysis~\label{sec:analysis}}

In this analysis, we consider a sterile neutrino model parameterized by one mass-squared difference, $\Delta m ^2_{41}$, and one mixing angle, $\sin^2(\theta_{24})$.
For each hypothesis point on a grid of $\Delta m^2_{41}$ from $10^{-2}~{\rm eV}^2$ to $10^{2}~{\rm eV}^2$ and $\sin^2(2 \theta_{24})$ from $10^{-3}$ to 1, the neutrino flux incident on the detector is calculated using the four-flavor formalism.

The neutrino flux includes contributions from both atmospheric and astrophysical neutrinos.
The conventional atmospheric $\nu_\mu$ and $\bar{\nu}_\mu$ flux is produced by the decay of pions and kaons and is calculated using the MCEq cascade equation solver~\cite{Fedynitch:2015zma,Fedynitch:2012fs}.
The hadronic interactions are modeled with {\tt Sibyll2.3c}~\cite{Riehn:2017mfm}. The primary cosmic-ray spectrum is a three-population model~\cite{Gaisser:2011cc,Hillas:2006ms}, in which each population contains five groups of nuclei.
The zenith-dependent seasonal atmospheric density profile, through which the cascade develops, is determined using data from the Atmospheric Infrared Sounder (AIRS) satellite~\cite{AIRS}.  
The prompt $\nu_\mu$ component from the decay of charmed mesons is implemented as in Ref.~\cite{Bhattacharya:2016jce}.
The astrophysical neutrino flux is assumed to have equal parts of each neutrino flavor and to be symmetric in neutrinos and antineutrinos~\cite{Palladino:2015vna,Arguelles:2015dca,Bustamante:2015waa}; be isotropically distributed; and have a single power-law energy spectrum consistent with previous IceCube measurements~\cite{Schneider:2019ayi}.
These fluxes are subject to a suite of systematic uncertainties, summarized in the following section.

For each sterile neutrino hypothesis, the atmospheric and astrophysical neutrino fluxes are propagated through the Earth using the \texttt{nuSQuIDS} neutrino evolution code~\cite{Delgado:2014kpa,nusquids}.
This accounts for both coherent and non-coherent interactions~\cite{GonzalezGarcia:2005xw}; namely charged-current, neutral-current, and Glashow resonance interactions~\cite{Glashow:1960zz}, as well as tau-neutrino regeneration~\cite{Halzen:1998be}.
We use the CSMS~\cite{CooperSarkar:2011pa} neutrino-nucleon cross section to describe both interactions during neutrino propagation and near the detector.
This requires as an input the Earth density profile, which we parameterize via the spherically symmetric PREM model~\cite{Dziewonski:1981xy}.
Using the above, we obtain a prediction for the up-going $\nu_\mu$ flux at the detector for each physics parameter point.
These fluxes are used to weight detector Monte Carlo (MC) event sets, with effective livetime $\geq 50\times$ the sample size.

We account for systematic uncertainties by means of nuisance parameters, which reweight the MC by applying continuous parameterizations of the effects discussed in the following section.
We then compare the data to expectation using a modified version of the Poisson likelihood to account for MC statistical uncertainty~\cite{Arguelles:2019izp}.
For our frequentist analysis, the likelihood is profiled over the eighteen nuisance parameters to construct a test statistic.
Frequentist contours are constructed using Wilks' theorem~\cite{wilks1938}, validated at an array of parameter points using MC ensembles and the Feldman-Cousins~\cite{Feldman:1997qc} procedure.
A Bayesian hypothesis test is also performed, by means of comparing the model evidences~\cite{RevModPhys.83.943} with respect to the no sterile neutrino hypothesis.
The model evidences, as a function of sterile neutrino parameters, are computed by integrating the likelihood over the nuisance parameters using \texttt{MultiNest}~\cite{Feroz:2008xx}. 
These two statistical approaches are complementary: the Bayesian approach conveys the likelihood of the model given observed data and prior knowledge, whereas the frequentist approach yields intervals that are likely to contain the true model parameters for repeated experiments, enabling direct comparison with previous publications.

\section{Systematic Uncertainties}\label{sec:Systematics}

Dominant sources of uncertainty derive from the shape and normalization of astrophysical and atmospheric neutrino fluxes; the bulk properties of the South Pole ice; the local response of the IceCube DOMs; and neutrino interaction cross sections.
Other uncertainties, such as the Earth density profile, neutrino interactions in the rock/ice transition region, prompt neutrino flux, and $\nu_\mu$/$\bar{\nu}_\mu$ astrophysical ratio were investigated but established as negligible relative to statistical uncertainty.

{\bf Atmospheric Neutrino Flux:}
In the relevant energy range the spectrum of cosmic-ray primaries follows approximately an $E^{-2.65}$ energy (E) dependence.
To account for the uncertainty in the cosmic-ray spectral index, we apply a spectral shift $\Delta \gamma$ with an uncertainty of 0.03 pivoting at 2.2\,TeV~\cite{Karelin:2011zz,Bartoli:2015fhw,Yoon:2017qjx,Alfaro:2017cwx}. 
The meson production uncertainty in the interaction between the primary cosmic ray and air and in subsequent hadronic interactions is described through the Barr \textit{et al}. scheme~\cite{Barr:2006it}.
In this scheme, the uncertainty in the differential cross section for meson production is quantified in regions of primary proton energy $E_p$ and meson fractional momenta $x_{\rm lab}$.
The charged-kaon production yield carries the leading uncertainty.
We parameterize its production over three kinematic regions: $x_{\rm lab} < 0.1$ and $E_p > 30~{\rm GeV}$; $x_{\rm lab} \geq 0.1$ and $30~{\rm GeV} < E_p < 500~{\rm GeV}$; and $x_{\rm lab} > 0.1$ and $E_p \geq 500~{\rm GeV}$.
We include two collider-constrained nuisance parameters for each region, one for particles and one for antiparticles, which rescale the production cross section.
The highest-energy uncertainties are obtained through extrapolation, and both the scale and energy dependence have ascribed uncertainties.
Kaon energy losses by interaction with oxygen and nitrogen nuclei are accounted for via the total kaon-nucleus cross-sectional uncertainty~\cite{PhysRevD.85.074020}.
The charged-pion production and interaction uncertainties were studied and found negligible.
The atmospheric density profile is inferred from the zenith-dependent vertical temperature profile measured by the AIRS satellite.
To incorporate its uncertainty, showers are recomputed through randomly perturbed density models within the statistical and systematic uncertainties reported in the AIRS measurements.
Finally, the total conventional atmospheric $\nu_\mu$ flux carries an additional 40\% normalization uncertainty following Ref.~\cite{Fedynitch:2012fs}.

{\bf Astrophysical Neutrino Flux:} 
The central astrophysical model is a single power law with an equal normalization for all neutrino and antineutrino flavors at 100\,TeV of 0.787$\times$10$^{-18}$\,GeV$^{-1}$sr$^{-1}$s$^{-1}$cm$^{-2}$ and a spectral index of 2.5.
The Gaussian priors on the normalization and spectral index are correlated and selected to accommodate all IceCube astrophysical neutrino flux measurements to date~\cite{Aartsen:2013jdh,Aartsen:2014gkd,Aartsen:2016xlq,Aartsen:2018vez,Schneider:2019ayi,Aartsen:2020aqd}, with allowed spectral indices of $\gamma_{\rm astro}\sim 2.2 - 2.8$ at 68\% confidence level (C.L.). This represents a significant contribution to the total flux in the top two energy bins, depending strongly on the value of  $\gamma_{\rm astro}$.

{\bf Bulk Ice Model:}
The uncertainty associated with the measured scattering and absorption of the undisturbed glacial ice is implemented as described in Ref.~\cite{Aartsen:2019jcj}.
This treatment expresses the depth dependence of the ice optical properties using a Fourier decomposition. 
The covariance of the Fourier mode  coefficients are determined using LED flasher calibration data~\cite{Aartsen:2013rt}.
Only the six lowest modes contribute a sizeable shape difference in the reconstructed event distributions.
The effect of these modes is parameterized using two empirical energy-dependent basis functions.
The two associated amplitudes are incorporated as nuisance parameters with a correlated bivariate Gaussian prior.

{\bf DOM Response and Local Ice Effects:}
The ice in the immediate vicinity of the DOMs has optical properties distinct from the bulk ice between strings~\cite{Karle:1994eua}, caused by bubble formation during the refreezing process after their deployment.
This introduces uncertainties via two effects.
First, the global photon detection efficiency is impacted.
This is modeled by an efficiency correction with an effectively flat prior, ultimately constrained with a tight posterior through its effect on the overall energy scale.
Second, the bubble column influences the angular dependence of photon detection.
This is encoded in two parameters tuned to detailed optical simulations of bubble scattering near the DOM~\cite{Aartsen:2017nmd}, with only one having a substantial impact.

{\bf Neutrino Cross Section:}
The neutrino-nucleon cross section enters the analysis in two ways, influencing: 1) absorption during the neutrino propagation through the Earth~\cite{Gandhi:1995tf,Vincent:2017svp} and 2) the rates and inelasticities of interactions near the detector~\cite{Gandhi:1995tf,Connolly:2011vc,CooperSarkar:2011pa}.
The latter source of uncertainty was previously investigated in Refs.~\cite{delgado2015new,jones2015sterile} and found to be negligible.
The former is found to be non-negligible and is taken into account by separately parameterizing the change in neutrino absorption when the $\nu_\mu$ and $\bar{\nu}_\mu$ cross sections are scaled.
The priors on these parameters are fixed at the largest uncertainties in our energy range from Ref.~\cite{CooperSarkar:2011pa}, which are 3\% for $\nu_\mu$ and 7\% for $\bar{\nu}_\mu$.

\section{Results\label{sec:results}}

\begin{figure}[ht]  
 \begin{minipage}{0.49\textwidth}
   \includegraphics[width=\textwidth]{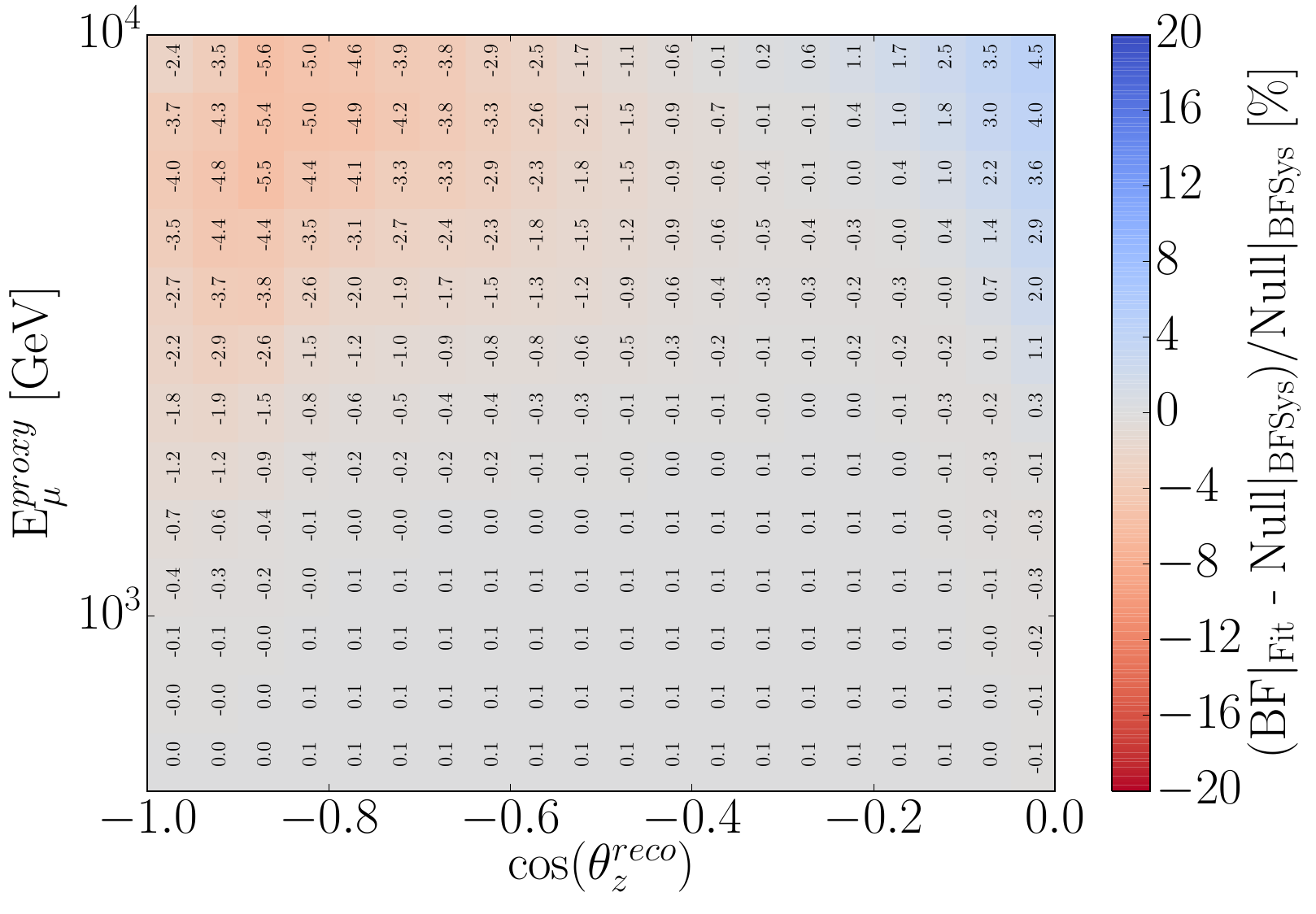}
  \end{minipage}
  \hfill
  \begin{minipage}{0.49\textwidth}
   \includegraphics[width=\textwidth]{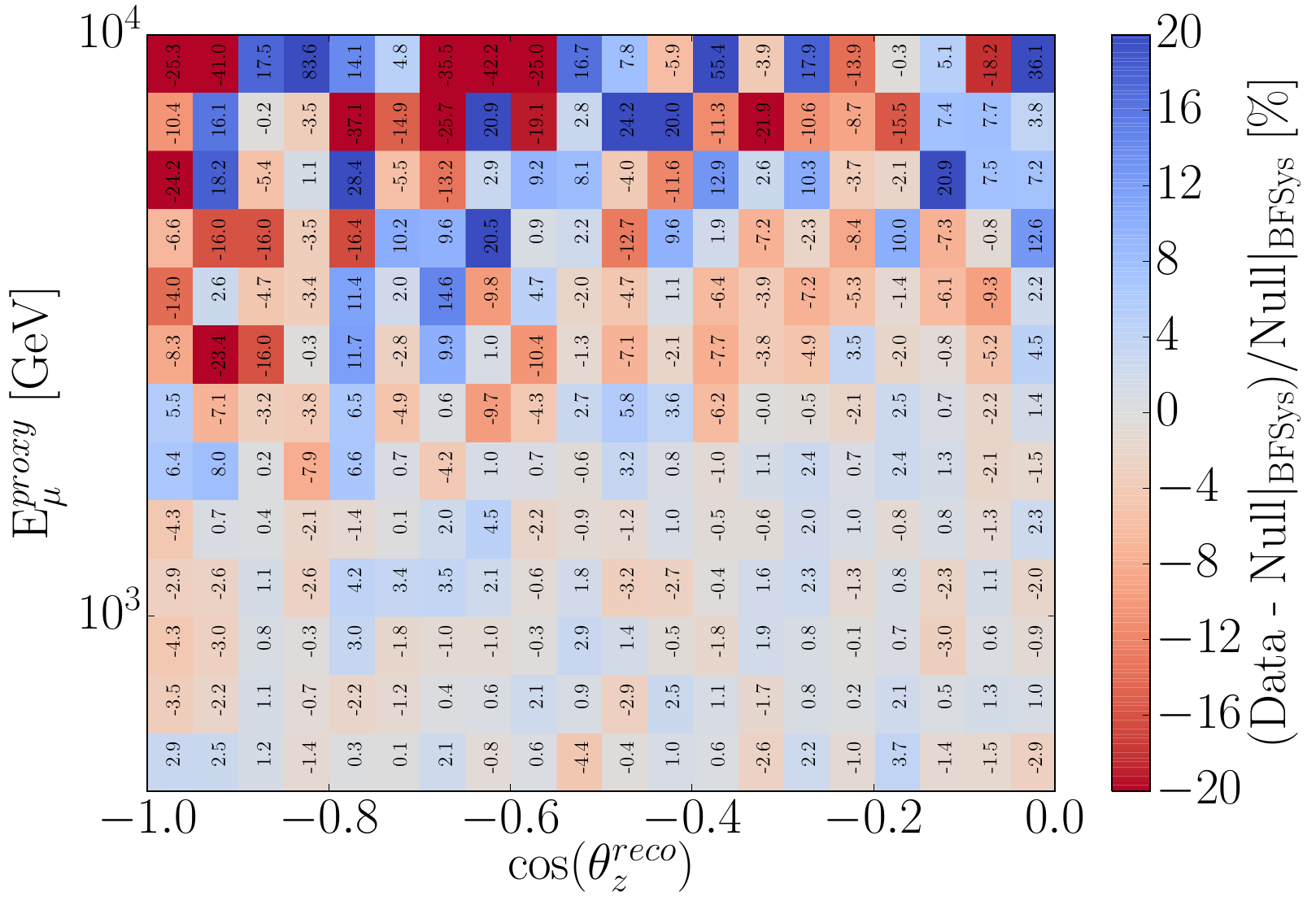}
  \end{minipage}
 \caption{\textbf{\textit{Best-fit signal shapes compared to data.}} Top: The signal shape at the best-fit point compared to the null hypothesis with the same nuisance parameters.
 Bottom:  data compared to the null hypothesis with the nuisance parameters held at the same values.}
  \label{fig::precut}
\end{figure}    

The frequentist analysis best-fit point is $\Delta m^2_{41}=4.5~{\rm eV}^2$ and $\sin^2(2\theta_{24})=0.10$.
At this point, the largest nuisance parameter pull was observed in the cosmic-ray spectral index, which shifted the cosmic-ray spectrum by 0.068 (2.3$\sigma$); the other nuisance parameter best-fit values are within one sigma of their respective central values and can be found in the accompanying Ref.~\cite{MEOWSPRD}.
Fig.~\ref{fig::precut} shows the signal shape at the best-fit point, given the best-fit nuisance parameters, as well as the pull between data and no sterile neutrino hypothesis, evaluated at those same nuisance parameters.
\begin{figure}[!h]  
    \includegraphics[width=0.90\columnwidth]{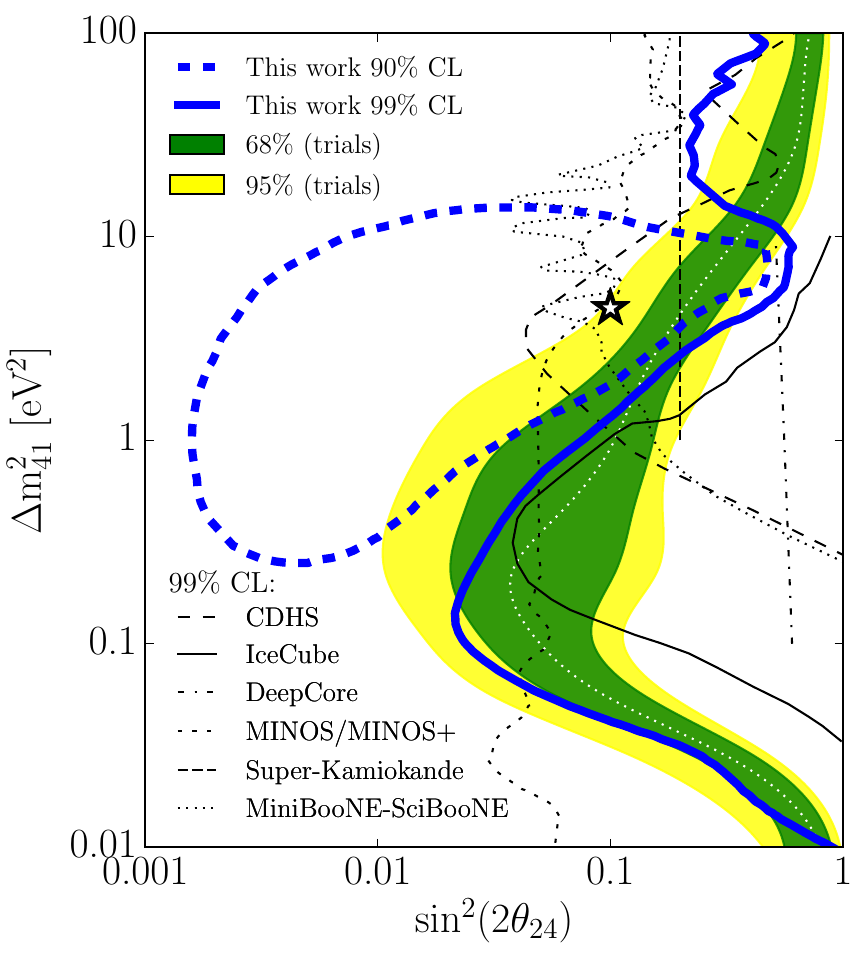}
    \caption{\textbf{\textit{Frequentist analysis result.}}
    The 90\% and 99\% C.L. contours, assuming Wilks' theorem, shown as dashed and solid bold blue lines respectively. 
    The green / yellow band shows the region where 68\% / 95\% of the pseudoexperiment 99\% C.L. observations lie; the dashed white line corresponds to the median.
    Other muon-neutrino disappearance measurements at 99\% C.L. are shown in black~\cite{TheIceCube:2016oqi,Aartsen:2017bap,MINOS:2016viw,Abe:2014gda,Mahn:2011ea,Stockdale:1984cg,Collin:2016aqd}; where results were not available at 99\% C.L., methods of Ref.~\cite{Diaz:2019fwt} were applied using public data releases. 
    Finally, the star marks the analysis best-fit point location.}
    \label{fig::freq_result}
\end{figure} 
Fig.~\ref{fig::freq_result} shows the 90\% and 99\% C.L. contours calculated according to Wilks' theorem with two degrees of freedom.
Sensitivity envelopes, illustrating symmetrically counted ensembles of 68\% and 95\% non-closed contours derived from 2,000 pseudoexperiments, are shown overlaid for the 99\% contour.
The IceCube 90\% C.L. preferred region is consistent with constraints from previous $\nu_\mu$ disappearance experiments, and the 99\% contour is stronger than other exclusion limits at values of $\Delta m^2$ up to 1\,eV$^2$.

\begin{figure}[t]  
    \includegraphics[width=1\columnwidth]{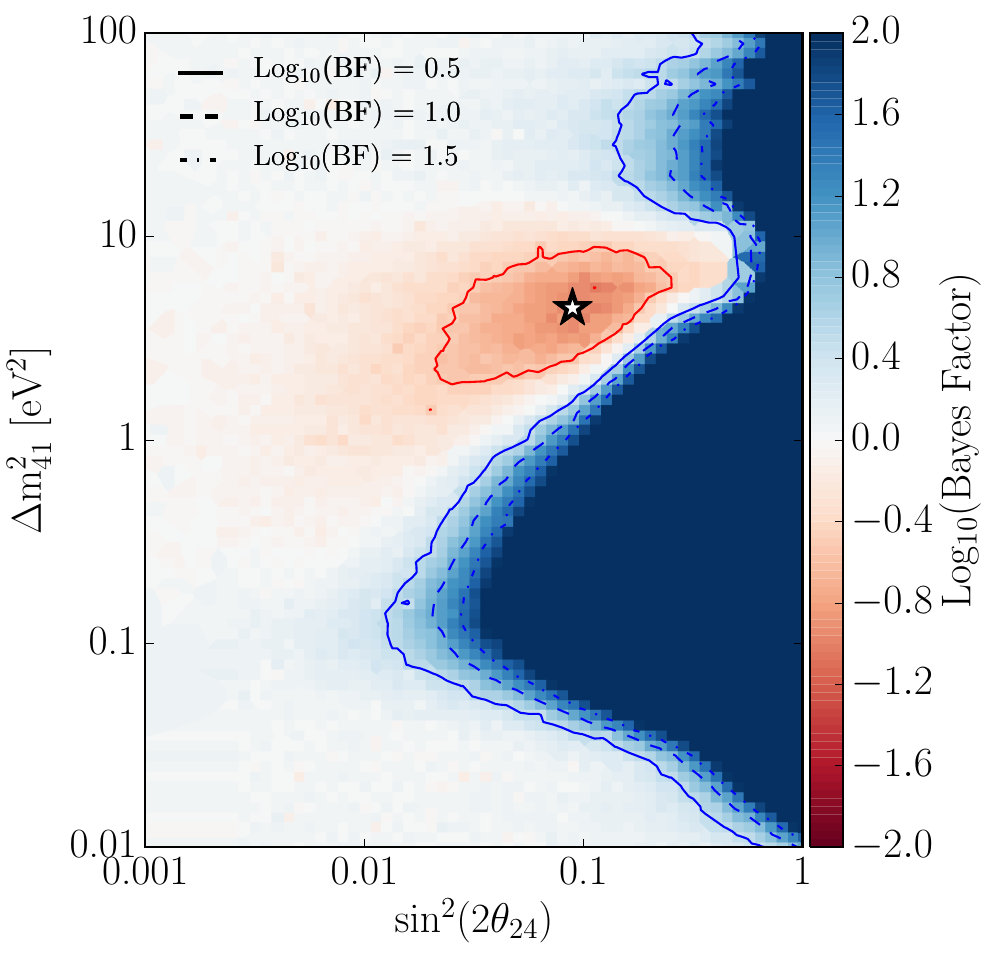}
    \caption{\textbf{\textit{Bayesian analysis result.}}
    The logarithm of the Bayes Factor~\cite{jeffreys1998theory} relative to the null hypothesis (color scale).
    Red indicates hypotheses preferred over the null hypothesis, while the blue indicates the null is preferred.
    Solid lines delineate likelihood ratios of 1 in 10 for \textit{a priori} equally likely hypotheses.
    The best-model location is shown at the white star with a $\log_{10}$(Bayes Factor) minimum of $-1.03$. }
    \label{fig::bayesian_result}
\end{figure}

Fig.~\ref{fig::bayesian_result} shows the corresponding Bayesian result, where the point-wise Bayes factor is calculated relative to the no sterile neutrino hypothesis.
The best-model location is at $\Delta m^2 \sim 4.5\,{\rm eV}^2$ and $\sin^2(2\theta_{24}) \sim 0.9$ and is strongly preferred, by a factor of 10.7, to the no sterile neutrino hypothesis.
Contours are drawn in logarithmic Bayes Factor steps of 0.5, quantifying strength of evidence~\cite{jeffreys1998theory}.

The best-fit point and inferred confidence regions are found to be robust under the removal of any one of the eight years of data, showing only minor changes in the contour position.
This is also the case for removal of any of the following group of uncertainties: neutrino cross sections, detector effects, atmospheric flux, and astrophysical flux. Details can be found in Ref.~\cite{MEOWSPRD}.
Furthermore, a similar best-fit point is obtained when fitting any one year of data independently, suggesting a small effect of physical or systematic rather than statistical origin.

The difference in likelihood to the null hypothesis is 4.94, corresponding to a p-value of 8\% against the null hypothesis.
The location of this point was found to be compatible with expectations based on simulated no sterile neutrino pseudoexperiments, which by definition produce closed contours at 90\%~C.L. in 10\% of trials.

In summary, we have studied 305,735 up-going atmospheric and astrophysical muon-neutrinos to search for evidence of eV-sterile neutrino signatures.
The best-fit point is consistent with the no sterile neutrino hypothesis at a p-value of 8\%.
Because of its unique statistical strength this result is expected to have a substantial impact on the global sterile neutrino landscape.

\section{Acknowledgements\label{sec:ack}}

\begin{acknowledgments}

The IceCube collaboration acknowledges the significant contributions to this manuscript from the Massachusetts Institute of Technology and University of Texas at Arlington groups.

We acknowledge the support from the following agencies: 
USA {\textendash} U.S. National Science Foundation-Office of Polar Programs,
U.S. National Science Foundation-Physics Division,
Wisconsin Alumni Research Foundation,
Center for High Throughput Computing (CHTC) at the University of Wisconsin-Madison,
Open Science Grid (OSG),
Extreme Science and Engineering Discovery Environment (XSEDE),
U.S. Department of Energy-National Energy Research Scientific Computing Center,
Particle astrophysics research computing center at the University of Maryland,
Institute for Cyber-Enabled Research at Michigan State University,
and Astroparticle physics computational facility at Marquette University;
Belgium {\textendash} Funds for Scientific Research (FRS-FNRS and FWO),
FWO Odysseus and Big Science programmes,
and Belgian Federal Science Policy Office (Belspo);
Germany {\textendash} Bundesministerium f{\"u}r Bildung und Forschung (BMBF),
Deutsche Forschungsgemeinschaft (DFG),
Helmholtz Alliance for Astroparticle Physics (HAP),
Initiative and Networking Fund of the Helmholtz Association,
Deutsches Elektronen Synchrotron (DESY),
and High Performance Computing cluster of the RWTH Aachen;
Sweden {\textendash} Swedish Research Council,
Swedish Polar Research Secretariat,
Swedish National Infrastructure for Computing (SNIC),
and Knut and Alice Wallenberg Foundation;
Australia {\textendash} Australian Research Council;
Canada {\textendash} Natural Sciences and Engineering Research Council of Canada,
Calcul Qu{\'e}bec, Compute Ontario, Canada Foundation for Innovation, WestGrid, and Compute Canada;
Denmark {\textendash} Villum Fonden, Danish National Research Foundation (DNRF), Carlsberg Foundation;
New Zealand {\textendash} Marsden Fund;
Japan {\textendash} Japan Society for Promotion of Science (JSPS)
and Institute for Global Prominent Research (IGPR) of Chiba University;
Korea {\textendash} National Research Foundation of Korea (NRF);
Switzerland {\textendash} Swiss National Science Foundation (SNSF);
United Kingdom {\textendash} Department of Physics, University of Oxford.
\end{acknowledgments}

\bibliography{meows_prl}

\begin{thebibliography}{102}%
\makeatletter
\providecommand \@ifxundefined [1]{%
 \@ifx{#1\undefined}
}%
\providecommand \@ifnum [1]{%
 \ifnum #1\expandafter \@firstoftwo
 \else \expandafter \@secondoftwo
 \fi
}%
\providecommand \@ifx [1]{%
 \ifx #1\expandafter \@firstoftwo
 \else \expandafter \@secondoftwo
 \fi
}%
\providecommand \natexlab [1]{#1}%
\providecommand \enquote  [1]{``#1''}%
\providecommand \bibnamefont  [1]{#1}%
\providecommand \bibfnamefont [1]{#1}%
\providecommand \citenamefont [1]{#1}%
\providecommand \href@noop [0]{\@secondoftwo}%
\providecommand \href [0]{\begingroup \@sanitize@url \@href}%
\providecommand \@href[1]{\@@startlink{#1}\@@href}%
\providecommand \@@href[1]{\endgroup#1\@@endlink}%
\providecommand \@sanitize@url [0]{\catcode `\\12\catcode `\$12\catcode
  `\&12\catcode `\#12\catcode `\^12\catcode `\_12\catcode `\%12\relax}%
\providecommand \@@startlink[1]{}%
\providecommand \@@endlink[0]{}%
\providecommand \url  [0]{\begingroup\@sanitize@url \@url }%
\providecommand \@url [1]{\endgroup\@href {#1}{\urlprefix }}%
\providecommand \urlprefix  [0]{URL }%
\providecommand \Eprint [0]{\href }%
\providecommand \doibase [0]{https://doi.org/}%
\providecommand \selectlanguage [0]{\@gobble}%
\providecommand \bibinfo  [0]{\@secondoftwo}%
\providecommand \bibfield  [0]{\@secondoftwo}%
\providecommand \translation [1]{[#1]}%
\providecommand \BibitemOpen [0]{}%
\providecommand \bibitemStop [0]{}%
\providecommand \bibitemNoStop [0]{.\EOS\space}%
\providecommand \EOS [0]{\spacefactor3000\relax}%
\providecommand \BibitemShut  [1]{\csname bibitem#1\endcsname}%
\let\auto@bib@innerbib\@empty
\bibitem [{\citenamefont {Tanabashi}\ \emph {et~al.}(2018)\citenamefont
  {Tanabashi} \emph {et~al.}}]{Tanabashi:2018oca}%
  \BibitemOpen
  \bibfield  {author} {\bibinfo {author} {\bibfnamefont {M.}~\bibnamefont
  {Tanabashi}} \emph {et~al.} (\bibinfo {collaboration} {Particle Data
  Group}),\ }\bibfield  {title} {\bibinfo {title} {{Review of Particle
  Physics}},\ }\href {https://doi.org/10.1103/PhysRevD.98.030001} {\bibfield
  {journal} {\bibinfo  {journal} {Phys. Rev.}\ }\textbf {\bibinfo {volume}
  {D98}},\ \bibinfo {pages} {030001} (\bibinfo {year} {2018})}\BibitemShut
  {NoStop}%
\bibitem [{\citenamefont {Esteban}\ \emph {et~al.}(2019)\citenamefont
  {Esteban}, \citenamefont {Gonzalez-Garcia}, \citenamefont
  {Hern\'andez-Cabezudo}, \citenamefont {Maltoni},\ and\ \citenamefont
  {Schwetz}}]{Esteban:2018azc}%
  \BibitemOpen
  \bibfield  {author} {\bibinfo {author} {\bibfnamefont {I.}~\bibnamefont
  {Esteban}}, \bibinfo {author} {\bibfnamefont {M.~C.}\ \bibnamefont
  {Gonzalez-Garcia}}, \bibinfo {author} {\bibfnamefont {A.}~\bibnamefont
  {Hern\'andez-Cabezudo}}, \bibinfo {author} {\bibfnamefont {M.}~\bibnamefont
  {Maltoni}},\ and\ \bibinfo {author} {\bibfnamefont {T.}~\bibnamefont
  {Schwetz}},\ }\bibfield  {title} {\bibinfo {title} {{Global analysis of
  three-flavour neutrino oscillations: synergies and tensions in the
  determination of $\theta_{23}$, $\delta_{CP}$, and the mass ordering}},\
  }\href {https://doi.org/10.1007/JHEP01(2019)106} {\bibfield  {journal}
  {\bibinfo  {journal} {JHEP}\ }\textbf {\bibinfo {volume} {01}},\ \bibinfo
  {pages} {106}},\ \Eprint {https://arxiv.org/abs/1811.05487} {arXiv:1811.05487
  [hep-ph]} \BibitemShut {NoStop}%
\bibitem [{\citenamefont {de~Salas}\ \emph {et~al.}(2018)\citenamefont
  {de~Salas}, \citenamefont {Forero}, \citenamefont {Ternes}, \citenamefont
  {Tortola},\ and\ \citenamefont {Valle}}]{deSalas:2017kay}%
  \BibitemOpen
  \bibfield  {author} {\bibinfo {author} {\bibfnamefont {P.~F.}\ \bibnamefont
  {de~Salas}}, \bibinfo {author} {\bibfnamefont {D.~V.}\ \bibnamefont
  {Forero}}, \bibinfo {author} {\bibfnamefont {C.~A.}\ \bibnamefont {Ternes}},
  \bibinfo {author} {\bibfnamefont {M.}~\bibnamefont {Tortola}},\ and\ \bibinfo
  {author} {\bibfnamefont {J.~W.~F.}\ \bibnamefont {Valle}},\ }\bibfield
  {title} {\bibinfo {title} {{Status of neutrino oscillations 2018: 3$\sigma$
  hint for normal mass ordering and improved $CP$ sensitivity}},\ }\href
  {https://doi.org/10.1016/j.physletb.2018.06.019} {\bibfield  {journal}
  {\bibinfo  {journal} {Phys. Lett.}\ }\textbf {\bibinfo {volume} {B782}},\
  \bibinfo {pages} {633} (\bibinfo {year} {2018})},\ \Eprint
  {https://arxiv.org/abs/1708.01186} {arXiv:1708.01186 [hep-ph]} \BibitemShut
  {NoStop}%
\bibitem [{\citenamefont {Capozzi}\ \emph {et~al.}(2016)\citenamefont
  {Capozzi}, \citenamefont {Lisi}, \citenamefont {Marrone}, \citenamefont
  {Montanino},\ and\ \citenamefont {Palazzo}}]{Capozzi:2016rtj}%
  \BibitemOpen
  \bibfield  {author} {\bibinfo {author} {\bibfnamefont {F.}~\bibnamefont
  {Capozzi}}, \bibinfo {author} {\bibfnamefont {E.}~\bibnamefont {Lisi}},
  \bibinfo {author} {\bibfnamefont {A.}~\bibnamefont {Marrone}}, \bibinfo
  {author} {\bibfnamefont {D.}~\bibnamefont {Montanino}},\ and\ \bibinfo
  {author} {\bibfnamefont {A.}~\bibnamefont {Palazzo}},\ }\bibfield  {title}
  {\bibinfo {title} {{Neutrino masses and mixings: Status of known and unknown
  $3\nu$ parameters}},\ }\href
  {https://doi.org/10.1016/j.nuclphysb.2016.02.016} {\bibfield  {journal}
  {\bibinfo  {journal} {Nucl. Phys.}\ }\textbf {\bibinfo {volume} {B908}},\
  \bibinfo {pages} {218} (\bibinfo {year} {2016})},\ \Eprint
  {https://arxiv.org/abs/1601.07777} {arXiv:1601.07777 [hep-ph]} \BibitemShut
  {NoStop}%
\bibitem [{\citenamefont {Athanassopoulos}\ \emph {et~al.}(1998)\citenamefont
  {Athanassopoulos} \emph {et~al.}}]{Athanassopoulos:1997pv}%
  \BibitemOpen
  \bibfield  {author} {\bibinfo {author} {\bibfnamefont {C.}~\bibnamefont
  {Athanassopoulos}} \emph {et~al.} (\bibinfo {collaboration} {LSND}),\
  }\bibfield  {title} {\bibinfo {title} {{Evidence for $\nu_\mu \to \nu_e$
  neutrino oscillations from LSND}},\ }\href
  {https://doi.org/10.1103/PhysRevLett.81.1774} {\bibfield  {journal} {\bibinfo
   {journal} {Phys. Rev. Lett.}\ }\textbf {\bibinfo {volume} {81}},\ \bibinfo
  {pages} {1774} (\bibinfo {year} {1998})},\ \Eprint
  {https://arxiv.org/abs/nucl-ex/9709006} {arXiv:nucl-ex/9709006 [nucl-ex]}
  \BibitemShut {NoStop}%
\bibitem [{\citenamefont {Aguilar-Arevalo}\ \emph {et~al.}(2018)\citenamefont
  {Aguilar-Arevalo} \emph {et~al.}}]{Aguilar-Arevalo:2018gpe}%
  \BibitemOpen
  \bibfield  {author} {\bibinfo {author} {\bibfnamefont {A.~A.}\ \bibnamefont
  {Aguilar-Arevalo}} \emph {et~al.} (\bibinfo {collaboration} {MiniBooNE}),\
  }\bibfield  {title} {\bibinfo {title} {{Significant Excess of ElectronLike
  Events in the MiniBooNE Short-Baseline Neutrino Experiment}},\ }\href
  {https://doi.org/10.1103/PhysRevLett.121.221801} {\bibfield  {journal}
  {\bibinfo  {journal} {Phys. Rev. Lett.}\ }\textbf {\bibinfo {volume} {121}},\
  \bibinfo {pages} {221801} (\bibinfo {year} {2018})},\ \Eprint
  {https://arxiv.org/abs/1805.12028} {arXiv:1805.12028 [hep-ex]} \BibitemShut
  {NoStop}%
\bibitem [{\citenamefont {Abazajian}\ \emph {et~al.}(2012)\citenamefont
  {Abazajian} \emph {et~al.}}]{Abazajian:2012ys}%
  \BibitemOpen
  \bibfield  {author} {\bibinfo {author} {\bibfnamefont {K.~N.}\ \bibnamefont
  {Abazajian}} \emph {et~al.},\ }\bibfield  {title} {\bibinfo {title} {{Light
  Sterile Neutrinos: A White Paper}},\ }\href@noop {} {\  (\bibinfo {year}
  {2012})},\ \Eprint {https://arxiv.org/abs/1204.5379} {arXiv:1204.5379
  [hep-ph]} \BibitemShut {NoStop}%
\bibitem [{\citenamefont {Schael}\ \emph {et~al.}(2006)\citenamefont {Schael}
  \emph {et~al.}}]{ALEPH:2005ab}%
  \BibitemOpen
  \bibfield  {author} {\bibinfo {author} {\bibfnamefont {S.}~\bibnamefont
  {Schael}} \emph {et~al.} (\bibinfo {collaboration} {ALEPH, DELPHI, L3, OPAL,
  SLD, LEP Electroweak Working Group, SLD Electroweak Group, SLD Heavy Flavour
  Group}),\ }\bibfield  {title} {\bibinfo {title} {{Precision electroweak
  measurements on the $Z$ resonance}},\ }\href
  {https://doi.org/10.1016/j.physrep.2005.12.006} {\bibfield  {journal}
  {\bibinfo  {journal} {Phys. Rept.}\ }\textbf {\bibinfo {volume} {427}},\
  \bibinfo {pages} {257} (\bibinfo {year} {2006})},\ \Eprint
  {https://arxiv.org/abs/hep-ex/0509008} {arXiv:hep-ex/0509008 [hep-ex]}
  \BibitemShut {NoStop}%
\bibitem [{\citenamefont {Akhmedov}(1988)}]{Akhmedov:1988kd}%
  \BibitemOpen
  \bibfield  {author} {\bibinfo {author} {\bibfnamefont {E.~K.}\ \bibnamefont
  {Akhmedov}},\ }\bibfield  {title} {\bibinfo {title} {{Neutrino oscillations
  in inhomogeneous matter. (In Russian)}},\ }\href@noop {} {\bibfield
  {journal} {\bibinfo  {journal} {Sov. J. Nucl. Phys.}\ }\textbf {\bibinfo
  {volume} {47}},\ \bibinfo {pages} {301} (\bibinfo {year} {1988})},\ \bibinfo
  {note} {[Yad. Fiz.47,475(1988)]}\BibitemShut {NoStop}%
\bibitem [{\citenamefont {Krastev}\ and\ \citenamefont
  {Smirnov}(1989)}]{Krastev:1989ix}%
  \BibitemOpen
  \bibfield  {author} {\bibinfo {author} {\bibfnamefont {P.~I.}\ \bibnamefont
  {Krastev}}\ and\ \bibinfo {author} {\bibfnamefont {A.~{\relax Yu}.}\
  \bibnamefont {Smirnov}},\ }\bibfield  {title} {\bibinfo {title} {{Parametric
  Effects in Neutrino Oscillations}},\ }\href
  {https://doi.org/10.1016/0370-2693(89)91206-9} {\bibfield  {journal}
  {\bibinfo  {journal} {Phys. Lett.}\ }\textbf {\bibinfo {volume} {B226}},\
  \bibinfo {pages} {341} (\bibinfo {year} {1989})}\BibitemShut {NoStop}%
\bibitem [{\citenamefont {Chizhov}\ \emph {et~al.}(1998)\citenamefont
  {Chizhov}, \citenamefont {Maris},\ and\ \citenamefont
  {Petcov}}]{Chizhov:1998ug}%
  \BibitemOpen
  \bibfield  {author} {\bibinfo {author} {\bibfnamefont {M.}~\bibnamefont
  {Chizhov}}, \bibinfo {author} {\bibfnamefont {M.}~\bibnamefont {Maris}},\
  and\ \bibinfo {author} {\bibfnamefont {S.~T.}\ \bibnamefont {Petcov}},\
  }\bibfield  {title} {\bibinfo {title} {{On the oscillation length resonance
  in the transitions of solar and atmospheric neutrinos crossing the earth
  core}},\ }\href@noop {} {\  (\bibinfo {year} {1998})},\ \Eprint
  {https://arxiv.org/abs/hep-ph/9810501} {arXiv:hep-ph/9810501 [hep-ph]}
  \BibitemShut {NoStop}%
\bibitem [{\citenamefont {Chizhov}\ and\ \citenamefont
  {Petcov}(1999)}]{Chizhov:1999az}%
  \BibitemOpen
  \bibfield  {author} {\bibinfo {author} {\bibfnamefont {M.~V.}\ \bibnamefont
  {Chizhov}}\ and\ \bibinfo {author} {\bibfnamefont {S.~T.}\ \bibnamefont
  {Petcov}},\ }\bibfield  {title} {\bibinfo {title} {{New conditions for a
  total neutrino conversion in a medium}},\ }\href
  {https://doi.org/10.1103/PhysRevLett.83.1096} {\bibfield  {journal} {\bibinfo
   {journal} {Phys. Rev. Lett.}\ }\textbf {\bibinfo {volume} {83}},\ \bibinfo
  {pages} {1096} (\bibinfo {year} {1999})},\ \Eprint
  {https://arxiv.org/abs/hep-ph/9903399} {arXiv:hep-ph/9903399 [hep-ph]}
  \BibitemShut {NoStop}%
\bibitem [{\citenamefont {Akhmedov}\ and\ \citenamefont
  {Smirnov}(2000)}]{Akhmedov:1999va}%
  \BibitemOpen
  \bibfield  {author} {\bibinfo {author} {\bibfnamefont {E.~K.}\ \bibnamefont
  {Akhmedov}}\ and\ \bibinfo {author} {\bibfnamefont {A.~{\relax Yu}.}\
  \bibnamefont {Smirnov}},\ }\bibfield  {title} {\bibinfo {title} {{Comment on
  `New conditions for a total neutrino conversion in a medium'}},\ }\href
  {https://doi.org/10.1103/PhysRevLett.85.3978} {\bibfield  {journal} {\bibinfo
   {journal} {Phys. Rev. Lett.}\ }\textbf {\bibinfo {volume} {85}},\ \bibinfo
  {pages} {3978} (\bibinfo {year} {2000})},\ \Eprint
  {https://arxiv.org/abs/hep-ph/9910433} {arXiv:hep-ph/9910433 [hep-ph]}
  \BibitemShut {NoStop}%
\bibitem [{\citenamefont {Nunokawa}\ \emph {et~al.}(2003)\citenamefont
  {Nunokawa}, \citenamefont {Peres},\ and\ \citenamefont
  {Zukanovich~Funchal}}]{Nunokawa:2003ep}%
  \BibitemOpen
  \bibfield  {author} {\bibinfo {author} {\bibfnamefont {H.}~\bibnamefont
  {Nunokawa}}, \bibinfo {author} {\bibfnamefont {O.~L.~G.}\ \bibnamefont
  {Peres}},\ and\ \bibinfo {author} {\bibfnamefont {R.}~\bibnamefont
  {Zukanovich~Funchal}},\ }\bibfield  {title} {\bibinfo {title} {{Probing the
  LSND mass scale and four neutrino scenarios with a neutrino telescope}},\
  }\href {https://doi.org/10.1016/S0370-2693(03)00603-8} {\bibfield  {journal}
  {\bibinfo  {journal} {Phys. Lett.}\ }\textbf {\bibinfo {volume} {B562}},\
  \bibinfo {pages} {279} (\bibinfo {year} {2003})},\ \Eprint
  {https://arxiv.org/abs/hep-ph/0302039} {arXiv:hep-ph/0302039 [hep-ph]}
  \BibitemShut {NoStop}%
\bibitem [{\citenamefont {Choubey}(2007)}]{Choubey:2007ji}%
  \BibitemOpen
  \bibfield  {author} {\bibinfo {author} {\bibfnamefont {S.}~\bibnamefont
  {Choubey}},\ }\bibfield  {title} {\bibinfo {title} {{Signature of sterile
  species in atmospheric neutrino data at neutrino telescopes}},\ }\href
  {https://doi.org/10.1088/1126-6708/2007/12/014} {\bibfield  {journal}
  {\bibinfo  {journal} {JHEP}\ }\textbf {\bibinfo {volume} {12}},\ \bibinfo
  {pages} {014}},\ \Eprint {https://arxiv.org/abs/0709.1937} {arXiv:0709.1937
  [hep-ph]} \BibitemShut {NoStop}%
\bibitem [{\citenamefont {Barger}\ \emph {et~al.}(2012)\citenamefont {Barger},
  \citenamefont {Gao},\ and\ \citenamefont {Marfatia}}]{Barger:2011rc}%
  \BibitemOpen
  \bibfield  {author} {\bibinfo {author} {\bibfnamefont {V.}~\bibnamefont
  {Barger}}, \bibinfo {author} {\bibfnamefont {Y.}~\bibnamefont {Gao}},\ and\
  \bibinfo {author} {\bibfnamefont {D.}~\bibnamefont {Marfatia}},\ }\bibfield
  {title} {\bibinfo {title} {{Is there evidence for sterile neutrinos in
  IceCube data?}},\ }\href {https://doi.org/10.1103/PhysRevD.85.011302}
  {\bibfield  {journal} {\bibinfo  {journal} {Phys. Rev.}\ }\textbf {\bibinfo
  {volume} {D85}},\ \bibinfo {pages} {011302} (\bibinfo {year} {2012})},\
  \Eprint {https://arxiv.org/abs/1109.5748} {arXiv:1109.5748 [hep-ph]}
  \BibitemShut {NoStop}%
\bibitem [{\citenamefont {Esmaili}\ \emph {et~al.}(2012)\citenamefont
  {Esmaili}, \citenamefont {Halzen},\ and\ \citenamefont
  {Peres}}]{Esmaili:2012nz}%
  \BibitemOpen
  \bibfield  {author} {\bibinfo {author} {\bibfnamefont {A.}~\bibnamefont
  {Esmaili}}, \bibinfo {author} {\bibfnamefont {F.}~\bibnamefont {Halzen}},\
  and\ \bibinfo {author} {\bibfnamefont {O.~L.~G.}\ \bibnamefont {Peres}},\
  }\bibfield  {title} {\bibinfo {title} {{Constraining Sterile Neutrinos with
  AMANDA and IceCube Atmospheric Neutrino Data}},\ }\href
  {https://doi.org/10.1088/1475-7516/2012/11/041} {\bibfield  {journal}
  {\bibinfo  {journal} {JCAP}\ }\textbf {\bibinfo {volume} {1211}},\ \bibinfo
  {pages} {041}},\ \Eprint {https://arxiv.org/abs/1206.6903} {arXiv:1206.6903
  [hep-ph]} \BibitemShut {NoStop}%
\bibitem [{\citenamefont {Esmaili}\ and\ \citenamefont
  {Smirnov}(2013)}]{Esmaili:2013vza}%
  \BibitemOpen
  \bibfield  {author} {\bibinfo {author} {\bibfnamefont {A.}~\bibnamefont
  {Esmaili}}\ and\ \bibinfo {author} {\bibfnamefont {A.~{\relax Yu}.}\
  \bibnamefont {Smirnov}},\ }\bibfield  {title} {\bibinfo {title} {{Restricting
  the LSND and MiniBooNE sterile neutrinos with the IceCube atmospheric
  neutrino data}},\ }\href {https://doi.org/10.1007/JHEP12(2013)014} {\bibfield
   {journal} {\bibinfo  {journal} {JHEP}\ }\textbf {\bibinfo {volume} {12}},\
  \bibinfo {pages} {014}},\ \Eprint {https://arxiv.org/abs/1307.6824}
  {arXiv:1307.6824 [hep-ph]} \BibitemShut {NoStop}%
\bibitem [{\citenamefont {Lindner}\ \emph {et~al.}(2016)\citenamefont
  {Lindner}, \citenamefont {Rodejohann},\ and\ \citenamefont
  {Xu}}]{Lindner:2015iaa}%
  \BibitemOpen
  \bibfield  {author} {\bibinfo {author} {\bibfnamefont {M.}~\bibnamefont
  {Lindner}}, \bibinfo {author} {\bibfnamefont {W.}~\bibnamefont
  {Rodejohann}},\ and\ \bibinfo {author} {\bibfnamefont {X.-J.}\ \bibnamefont
  {Xu}},\ }\bibfield  {title} {\bibinfo {title} {{Sterile neutrinos in the
  light of IceCube}},\ }\href {https://doi.org/10.1007/JHEP01(2016)124}
  {\bibfield  {journal} {\bibinfo  {journal} {JHEP}\ }\textbf {\bibinfo
  {volume} {01}},\ \bibinfo {pages} {124}},\ \Eprint
  {https://arxiv.org/abs/1510.00666} {arXiv:1510.00666 [hep-ph]} \BibitemShut
  {NoStop}%
\bibitem [{\citenamefont {Diaz}\ \emph {et~al.}(2019)\citenamefont {Diaz},
  \citenamefont {Argüelles}, \citenamefont {Collin}, \citenamefont {Conrad},\
  and\ \citenamefont {Shaevitz}}]{Diaz:2019fwt}%
  \BibitemOpen
  \bibfield  {author} {\bibinfo {author} {\bibfnamefont {A.}~\bibnamefont
  {Diaz}}, \bibinfo {author} {\bibfnamefont {C.~A.}\ \bibnamefont
  {Argüelles}}, \bibinfo {author} {\bibfnamefont {G.~H.}\ \bibnamefont
  {Collin}}, \bibinfo {author} {\bibfnamefont {J.~M.}\ \bibnamefont {Conrad}},\
  and\ \bibinfo {author} {\bibfnamefont {M.~H.}\ \bibnamefont {Shaevitz}},\
  }\bibfield  {title} {\bibinfo {title} {{Where Are We With Light Sterile
  Neutrinos?}},\ }\href@noop {} {\  (\bibinfo {year} {2019})},\ \Eprint
  {https://arxiv.org/abs/1906.00045} {arXiv:1906.00045 [hep-ex]} \BibitemShut
  {NoStop}%
\bibitem [{\citenamefont {Dydak}\ \emph {et~al.}(1984)\citenamefont {Dydak}
  \emph {et~al.}}]{Dydak:1983zq}%
  \BibitemOpen
  \bibfield  {author} {\bibinfo {author} {\bibfnamefont {F.}~\bibnamefont
  {Dydak}} \emph {et~al.},\ }\bibfield  {title} {\bibinfo {title} {{A Search
  for Muon-neutrino Oscillations in the Delta m**2 Range 0.3-${\rm eV}^2$ to
  90-${\rm eV}^2$}},\ }\href {https://doi.org/10.1016/0370-2693(84)90688-9}
  {\bibfield  {journal} {\bibinfo  {journal} {Phys. Lett.}\ }\textbf {\bibinfo
  {volume} {134B}},\ \bibinfo {pages} {281} (\bibinfo {year}
  {1984})}\BibitemShut {NoStop}%
\bibitem [{\citenamefont {Stockdale}\ \emph {et~al.}(1984)\citenamefont
  {Stockdale} \emph {et~al.}}]{Stockdale:1984cg}%
  \BibitemOpen
  \bibfield  {author} {\bibinfo {author} {\bibfnamefont {I.~E.}\ \bibnamefont
  {Stockdale}} \emph {et~al.},\ }\bibfield  {title} {\bibinfo {title} {{Limits
  on Muon Neutrino Oscillations in the Mass Range 55-eV**2 < Delta m**2 <
  800-eV**2}},\ }\href {https://doi.org/10.1103/PhysRevLett.52.1384} {\bibfield
   {journal} {\bibinfo  {journal} {Phys. Rev. Lett.}\ }\textbf {\bibinfo
  {volume} {52}},\ \bibinfo {pages} {1384} (\bibinfo {year}
  {1984})}\BibitemShut {NoStop}%
\bibitem [{\citenamefont {Mahn}\ \emph {et~al.}(2012)\citenamefont {Mahn} \emph
  {et~al.}}]{Mahn:2011ea}%
  \BibitemOpen
  \bibfield  {author} {\bibinfo {author} {\bibfnamefont {K.~B.~M.}\
  \bibnamefont {Mahn}} \emph {et~al.} (\bibinfo {collaboration} {SciBooNE,
  MiniBooNE}),\ }\bibfield  {title} {\bibinfo {title} {{Dual baseline search
  for muon neutrino disappearance at $0.5 {\rm eV}^2 < \Delta m^2 < 40 {\rm
  eV}^2$}},\ }\href {https://doi.org/10.1103/PhysRevD.85.032007} {\bibfield
  {journal} {\bibinfo  {journal} {Phys. Rev.}\ }\textbf {\bibinfo {volume}
  {D85}},\ \bibinfo {pages} {032007} (\bibinfo {year} {2012})},\ \Eprint
  {https://arxiv.org/abs/1106.5685} {arXiv:1106.5685 [hep-ex]} \BibitemShut
  {NoStop}%
\bibitem [{\citenamefont {Abe}\ \emph {et~al.}(2015)\citenamefont {Abe} \emph
  {et~al.}}]{Abe:2014gda}%
  \BibitemOpen
  \bibfield  {author} {\bibinfo {author} {\bibfnamefont {K.}~\bibnamefont
  {Abe}} \emph {et~al.} (\bibinfo {collaboration} {Super-Kamiokande}),\
  }\bibfield  {title} {\bibinfo {title} {{Limits on sterile neutrino mixing
  using atmospheric neutrinos in Super-Kamiokande}},\ }\href
  {https://doi.org/10.1103/PhysRevD.91.052019} {\bibfield  {journal} {\bibinfo
  {journal} {Phys. Rev.}\ }\textbf {\bibinfo {volume} {D91}},\ \bibinfo {pages}
  {052019} (\bibinfo {year} {2015})},\ \Eprint
  {https://arxiv.org/abs/1410.2008} {arXiv:1410.2008 [hep-ex]} \BibitemShut
  {NoStop}%
\bibitem [{\citenamefont {Adamson}\ \emph {et~al.}(2016)\citenamefont {Adamson}
  \emph {et~al.}}]{MINOS:2016viw}%
  \BibitemOpen
  \bibfield  {author} {\bibinfo {author} {\bibfnamefont {P.}~\bibnamefont
  {Adamson}} \emph {et~al.} (\bibinfo {collaboration} {MINOS}),\ }\bibfield
  {title} {\bibinfo {title} {{Search for Sterile Neutrinos Mixing with Muon
  Neutrinos in MINOS}},\ }\href
  {https://doi.org/10.1103/PhysRevLett.117.151803} {\bibfield  {journal}
  {\bibinfo  {journal} {Phys. Rev. Lett.}\ }\textbf {\bibinfo {volume} {117}},\
  \bibinfo {pages} {151803} (\bibinfo {year} {2016})},\ \Eprint
  {https://arxiv.org/abs/1607.01176} {arXiv:1607.01176 [hep-ex]} \BibitemShut
  {NoStop}%
\bibitem [{\citenamefont {Aartsen}\ \emph
  {et~al.}(2016{\natexlab{a}})\citenamefont {Aartsen} \emph
  {et~al.}}]{TheIceCube:2016oqi}%
  \BibitemOpen
  \bibfield  {author} {\bibinfo {author} {\bibfnamefont {M.~G.}\ \bibnamefont
  {Aartsen}} \emph {et~al.} (\bibinfo {collaboration} {IceCube}),\ }\bibfield
  {title} {\bibinfo {title} {{Searches for Sterile Neutrinos with the IceCube
  Detector}},\ }\href {https://doi.org/10.1103/PhysRevLett.117.071801}
  {\bibfield  {journal} {\bibinfo  {journal} {Phys. Rev. Lett.}\ }\textbf
  {\bibinfo {volume} {117}},\ \bibinfo {pages} {071801} (\bibinfo {year}
  {2016}{\natexlab{a}})},\ \Eprint {https://arxiv.org/abs/1605.01990}
  {arXiv:1605.01990 [hep-ex]} \BibitemShut {NoStop}%
\bibitem [{\citenamefont {Aartsen}\ \emph
  {et~al.}(2017{\natexlab{a}})\citenamefont {Aartsen} \emph
  {et~al.}}]{Aartsen:2017bap}%
  \BibitemOpen
  \bibfield  {author} {\bibinfo {author} {\bibfnamefont {M.~G.}\ \bibnamefont
  {Aartsen}} \emph {et~al.} (\bibinfo {collaboration} {IceCube}),\ }\bibfield
  {title} {\bibinfo {title} {{Search for sterile neutrino mixing using three
  years of IceCube DeepCore data}},\ }\href
  {https://doi.org/10.1103/PhysRevD.95.112002} {\bibfield  {journal} {\bibinfo
  {journal} {Phys. Rev.}\ }\textbf {\bibinfo {volume} {D95}},\ \bibinfo {pages}
  {112002} (\bibinfo {year} {2017}{\natexlab{a}})},\ \Eprint
  {https://arxiv.org/abs/1702.05160} {arXiv:1702.05160 [hep-ex]} \BibitemShut
  {NoStop}%
\bibitem [{\citenamefont {Adamson}\ \emph {et~al.}(2019)\citenamefont {Adamson}
  \emph {et~al.}}]{Adamson:2017uda}%
  \BibitemOpen
  \bibfield  {author} {\bibinfo {author} {\bibfnamefont {P.}~\bibnamefont
  {Adamson}} \emph {et~al.} (\bibinfo {collaboration} {MINOS+}),\ }\bibfield
  {title} {\bibinfo {title} {{Search for sterile neutrinos in MINOS and MINOS+
  using a two-detector fit}},\ }\href
  {https://doi.org/10.1103/PhysRevLett.122.091803} {\bibfield  {journal}
  {\bibinfo  {journal} {Phys. Rev. Lett.}\ }\textbf {\bibinfo {volume} {122}},\
  \bibinfo {pages} {091803} (\bibinfo {year} {2019})},\ \Eprint
  {https://arxiv.org/abs/1710.06488} {arXiv:1710.06488 [hep-ex]} \BibitemShut
  {NoStop}%
\bibitem [{\citenamefont {Albert}\ \emph {et~al.}(2019)\citenamefont {Albert}
  \emph {et~al.}}]{Albert:2018mnz}%
  \BibitemOpen
  \bibfield  {author} {\bibinfo {author} {\bibfnamefont {A.}~\bibnamefont
  {Albert}} \emph {et~al.} (\bibinfo {collaboration} {ANTARES}),\ }\bibfield
  {title} {\bibinfo {title} {{Measuring the atmospheric neutrino oscillation
  parameters and constraining the 3+1 neutrino model with ten years of ANTARES
  data}},\ }\href {https://doi.org/10.1007/JHEP06(2019)113} {\bibfield
  {journal} {\bibinfo  {journal} {JHEP}\ }\textbf {\bibinfo {volume} {06}},\
  \bibinfo {pages} {113}},\ \Eprint {https://arxiv.org/abs/1812.08650}
  {arXiv:1812.08650 [hep-ex]} \BibitemShut {NoStop}%
\bibitem [{\citenamefont {Declais}\ \emph {et~al.}(1995)\citenamefont {Declais}
  \emph {et~al.}}]{Declais:1994su}%
  \BibitemOpen
  \bibfield  {author} {\bibinfo {author} {\bibfnamefont {Y.}~\bibnamefont
  {Declais}} \emph {et~al.},\ }\bibfield  {title} {\bibinfo {title} {{Search
  for neutrino oscillations at 15-meters, 40-meters, and 95-meters from a
  nuclear power reactor at Bugey}},\ }\href
  {https://doi.org/10.1016/0550-3213(94)00513-E} {\bibfield  {journal}
  {\bibinfo  {journal} {Nucl. Phys.}\ }\textbf {\bibinfo {volume} {B434}},\
  \bibinfo {pages} {503} (\bibinfo {year} {1995})}\BibitemShut {NoStop}%
\bibitem [{\citenamefont {Abdurashitov}\ \emph {et~al.}(2009)\citenamefont
  {Abdurashitov} \emph {et~al.}}]{Abdurashitov:2009tn}%
  \BibitemOpen
  \bibfield  {author} {\bibinfo {author} {\bibfnamefont {J.~N.}\ \bibnamefont
  {Abdurashitov}} \emph {et~al.} (\bibinfo {collaboration} {SAGE}),\ }\bibfield
   {title} {\bibinfo {title} {{Measurement of the solar neutrino capture rate
  with gallium metal. III: Results for the 2002--2007 data-taking period}},\
  }\href {https://doi.org/10.1103/PhysRevC.80.015807} {\bibfield  {journal}
  {\bibinfo  {journal} {Phys. Rev.}\ }\textbf {\bibinfo {volume} {C80}},\
  \bibinfo {pages} {015807} (\bibinfo {year} {2009})},\ \Eprint
  {https://arxiv.org/abs/0901.2200} {arXiv:0901.2200 [nucl-ex]} \BibitemShut
  {NoStop}%
\bibitem [{\citenamefont {Kaether}\ \emph {et~al.}(2010)\citenamefont
  {Kaether}, \citenamefont {Hampel}, \citenamefont {Heusser}, \citenamefont
  {Kiko},\ and\ \citenamefont {Kirsten}}]{Kaether:2010ag}%
  \BibitemOpen
  \bibfield  {author} {\bibinfo {author} {\bibfnamefont {F.}~\bibnamefont
  {Kaether}}, \bibinfo {author} {\bibfnamefont {W.}~\bibnamefont {Hampel}},
  \bibinfo {author} {\bibfnamefont {G.}~\bibnamefont {Heusser}}, \bibinfo
  {author} {\bibfnamefont {J.}~\bibnamefont {Kiko}},\ and\ \bibinfo {author}
  {\bibfnamefont {T.}~\bibnamefont {Kirsten}},\ }\bibfield  {title} {\bibinfo
  {title} {{Reanalysis of the GALLEX solar neutrino flux and source
  experiments}},\ }\href {https://doi.org/10.1016/j.physletb.2010.01.030}
  {\bibfield  {journal} {\bibinfo  {journal} {Phys. Lett.}\ }\textbf {\bibinfo
  {volume} {B685}},\ \bibinfo {pages} {47} (\bibinfo {year} {2010})},\ \Eprint
  {https://arxiv.org/abs/1001.2731} {arXiv:1001.2731 [hep-ex]} \BibitemShut
  {NoStop}%
\bibitem [{\citenamefont {Conrad}\ and\ \citenamefont
  {Shaevitz}(2012)}]{Conrad:2011ce}%
  \BibitemOpen
  \bibfield  {author} {\bibinfo {author} {\bibfnamefont {J.~M.}\ \bibnamefont
  {Conrad}}\ and\ \bibinfo {author} {\bibfnamefont {M.~H.}\ \bibnamefont
  {Shaevitz}},\ }\bibfield  {title} {\bibinfo {title} {{Limits on Electron
  Neutrino Disappearance from the KARMEN and LSND $\nu_e$ - Carbon Cross
  Section Data}},\ }\href {https://doi.org/10.1103/PhysRevD.85.013017}
  {\bibfield  {journal} {\bibinfo  {journal} {Phys. Rev.}\ }\textbf {\bibinfo
  {volume} {D85}},\ \bibinfo {pages} {013017} (\bibinfo {year} {2012})},\
  \Eprint {https://arxiv.org/abs/1106.5552} {arXiv:1106.5552 [hep-ex]}
  \BibitemShut {NoStop}%
\bibitem [{\citenamefont {Ko}\ \emph {et~al.}(2017)\citenamefont {Ko} \emph
  {et~al.}}]{Ko:2016owz}%
  \BibitemOpen
  \bibfield  {author} {\bibinfo {author} {\bibfnamefont {Y.~J.}\ \bibnamefont
  {Ko}} \emph {et~al.} (\bibinfo {collaboration} {NEOS}),\ }\bibfield  {title}
  {\bibinfo {title} {{Sterile Neutrino Search at the NEOS Experiment}},\ }\href
  {https://doi.org/10.1103/PhysRevLett.118.121802} {\bibfield  {journal}
  {\bibinfo  {journal} {Phys. Rev. Lett.}\ }\textbf {\bibinfo {volume} {118}},\
  \bibinfo {pages} {121802} (\bibinfo {year} {2017})},\ \Eprint
  {https://arxiv.org/abs/1610.05134} {arXiv:1610.05134 [hep-ex]} \BibitemShut
  {NoStop}%
\bibitem [{\citenamefont {An}\ \emph {et~al.}(2016)\citenamefont {An} \emph
  {et~al.}}]{An:2016luf}%
  \BibitemOpen
  \bibfield  {author} {\bibinfo {author} {\bibfnamefont {F.~P.}\ \bibnamefont
  {An}} \emph {et~al.} (\bibinfo {collaboration} {Daya Bay}),\ }\bibfield
  {title} {\bibinfo {title} {{Improved Search for a Light Sterile Neutrino with
  the Full Configuration of the Daya Bay Experiment}},\ }\href
  {https://doi.org/10.1103/PhysRevLett.117.151802} {\bibfield  {journal}
  {\bibinfo  {journal} {Phys. Rev. Lett.}\ }\textbf {\bibinfo {volume} {117}},\
  \bibinfo {pages} {151802} (\bibinfo {year} {2016})},\ \Eprint
  {https://arxiv.org/abs/1607.01174} {arXiv:1607.01174 [hep-ex]} \BibitemShut
  {NoStop}%
\bibitem [{\citenamefont {Alekseev}\ \emph {et~al.}(2018)\citenamefont
  {Alekseev} \emph {et~al.}}]{Alekseev:2018efk}%
  \BibitemOpen
  \bibfield  {author} {\bibinfo {author} {\bibfnamefont {I.}~\bibnamefont
  {Alekseev}} \emph {et~al.} (\bibinfo {collaboration} {DANSS}),\ }\bibfield
  {title} {\bibinfo {title} {{Search for sterile neutrinos at the DANSS
  experiment}},\ }\href {https://doi.org/10.1016/j.physletb.2018.10.038}
  {\bibfield  {journal} {\bibinfo  {journal} {Phys. Lett.}\ }\textbf {\bibinfo
  {volume} {B787}},\ \bibinfo {pages} {56} (\bibinfo {year} {2018})},\ \Eprint
  {https://arxiv.org/abs/1804.04046} {arXiv:1804.04046 [hep-ex]} \BibitemShut
  {NoStop}%
\bibitem [{\citenamefont {Ashenfelter}\ \emph {et~al.}(2018)\citenamefont
  {Ashenfelter} \emph {et~al.}}]{Ashenfelter:2018iov}%
  \BibitemOpen
  \bibfield  {author} {\bibinfo {author} {\bibfnamefont {J.}~\bibnamefont
  {Ashenfelter}} \emph {et~al.} (\bibinfo {collaboration} {PROSPECT}),\
  }\bibfield  {title} {\bibinfo {title} {{First search for short-baseline
  neutrino oscillations at HFIR with PROSPECT}},\ }\href
  {https://doi.org/10.1103/PhysRevLett.121.251802} {\bibfield  {journal}
  {\bibinfo  {journal} {Phys. Rev. Lett.}\ }\textbf {\bibinfo {volume} {121}},\
  \bibinfo {pages} {251802} (\bibinfo {year} {2018})},\ \Eprint
  {https://arxiv.org/abs/1806.02784} {arXiv:1806.02784 [hep-ex]} \BibitemShut
  {NoStop}%
\bibitem [{\citenamefont {Almazán}\ \emph {et~al.}(2018)\citenamefont
  {Almazán} \emph {et~al.}}]{Almazan:2018wln}%
  \BibitemOpen
  \bibfield  {author} {\bibinfo {author} {\bibfnamefont {H.}~\bibnamefont
  {Almazán}} \emph {et~al.} (\bibinfo {collaboration} {STEREO}),\ }\bibfield
  {title} {\bibinfo {title} {{Sterile Neutrino Constraints from the STEREO
  Experiment with 66 Days of Reactor-On Data}},\ }\href
  {https://doi.org/10.1103/PhysRevLett.121.161801} {\bibfield  {journal}
  {\bibinfo  {journal} {Phys. Rev. Lett.}\ }\textbf {\bibinfo {volume} {121}},\
  \bibinfo {pages} {161801} (\bibinfo {year} {2018})},\ \Eprint
  {https://arxiv.org/abs/1806.02096} {arXiv:1806.02096 [hep-ex]} \BibitemShut
  {NoStop}%
\bibitem [{\citenamefont {Armbruster}\ \emph {et~al.}(2002)\citenamefont
  {Armbruster} \emph {et~al.}}]{Armbruster:2002mp}%
  \BibitemOpen
  \bibfield  {author} {\bibinfo {author} {\bibfnamefont {B.}~\bibnamefont
  {Armbruster}} \emph {et~al.} (\bibinfo {collaboration} {KARMEN}),\ }\bibfield
   {title} {\bibinfo {title} {{Upper limits for neutrino oscillations
  muon-anti-neutrino $\rightarrow$ electron-anti-neutrino from muon decay at
  rest}},\ }\href {https://doi.org/10.1103/PhysRevD.65.112001} {\bibfield
  {journal} {\bibinfo  {journal} {Phys. Rev.}\ }\textbf {\bibinfo {volume}
  {D65}},\ \bibinfo {pages} {112001} (\bibinfo {year} {2002})},\ \Eprint
  {https://arxiv.org/abs/hep-ex/0203021} {arXiv:hep-ex/0203021 [hep-ex]}
  \BibitemShut {NoStop}%
\bibitem [{\citenamefont {Adamson}\ \emph {et~al.}(2009)\citenamefont {Adamson}
  \emph {et~al.}}]{Adamson:2008qj}%
  \BibitemOpen
  \bibfield  {author} {\bibinfo {author} {\bibfnamefont {P.}~\bibnamefont
  {Adamson}} \emph {et~al.} (\bibinfo {collaboration} {MiniBooNE, MINOS}),\
  }\bibfield  {title} {\bibinfo {title} {{First Measurement of $\nu_\mu$ and
  $\nu_e$ Events in an Off-Axis Horn-Focused Neutrino Beam}},\ }\href
  {https://doi.org/10.1103/PhysRevLett.102.211801} {\bibfield  {journal}
  {\bibinfo  {journal} {Phys. Rev. Lett.}\ }\textbf {\bibinfo {volume} {102}},\
  \bibinfo {pages} {211801} (\bibinfo {year} {2009})},\ \Eprint
  {https://arxiv.org/abs/0809.2447} {arXiv:0809.2447 [hep-ex]} \BibitemShut
  {NoStop}%
\bibitem [{\citenamefont {Astier}\ \emph {et~al.}(2003)\citenamefont {Astier}
  \emph {et~al.}}]{Astier:2003gs}%
  \BibitemOpen
  \bibfield  {author} {\bibinfo {author} {\bibfnamefont {P.}~\bibnamefont
  {Astier}} \emph {et~al.} (\bibinfo {collaboration} {NOMAD}),\ }\bibfield
  {title} {\bibinfo {title} {{Search for nu(mu) ---> nu(e) oscillations in the
  NOMAD experiment}},\ }\href {https://doi.org/10.1016/j.physletb.2003.07.029}
  {\bibfield  {journal} {\bibinfo  {journal} {Phys. Lett.}\ }\textbf {\bibinfo
  {volume} {B570}},\ \bibinfo {pages} {19} (\bibinfo {year} {2003})},\ \Eprint
  {https://arxiv.org/abs/hep-ex/0306037} {arXiv:hep-ex/0306037 [hep-ex]}
  \BibitemShut {NoStop}%
\bibitem [{\citenamefont {Gariazzo}\ \emph {et~al.}(2017)\citenamefont
  {Gariazzo}, \citenamefont {Giunti}, \citenamefont {Laveder},\ and\
  \citenamefont {Li}}]{Gariazzo:2017fdh}%
  \BibitemOpen
  \bibfield  {author} {\bibinfo {author} {\bibfnamefont {S.}~\bibnamefont
  {Gariazzo}}, \bibinfo {author} {\bibfnamefont {C.}~\bibnamefont {Giunti}},
  \bibinfo {author} {\bibfnamefont {M.}~\bibnamefont {Laveder}},\ and\ \bibinfo
  {author} {\bibfnamefont {Y.~F.}\ \bibnamefont {Li}},\ }\bibfield  {title}
  {\bibinfo {title} {{Updated Global 3+1 Analysis of Short-BaseLine Neutrino
  Oscillations}},\ }\href {https://doi.org/10.1007/JHEP06(2017)135} {\bibfield
  {journal} {\bibinfo  {journal} {JHEP}\ }\textbf {\bibinfo {volume} {06}},\
  \bibinfo {pages} {135}},\ \Eprint {https://arxiv.org/abs/1703.00860}
  {arXiv:1703.00860 [hep-ph]} \BibitemShut {NoStop}%
\bibitem [{\citenamefont {Dentler}\ \emph {et~al.}(2018)\citenamefont
  {Dentler}, \citenamefont {Hern\'andez-Cabezudo}, \citenamefont {Kopp},
  \citenamefont {Machado}, \citenamefont {Maltoni}, \citenamefont
  {Martinez-Soler},\ and\ \citenamefont {Schwetz}}]{Dentler:2018sju}%
  \BibitemOpen
  \bibfield  {author} {\bibinfo {author} {\bibfnamefont {M.}~\bibnamefont
  {Dentler}}, \bibinfo {author} {\bibfnamefont {A.}~\bibnamefont
  {Hern\'andez-Cabezudo}}, \bibinfo {author} {\bibfnamefont {J.}~\bibnamefont
  {Kopp}}, \bibinfo {author} {\bibfnamefont {P.~A.~N.}\ \bibnamefont
  {Machado}}, \bibinfo {author} {\bibfnamefont {M.}~\bibnamefont {Maltoni}},
  \bibinfo {author} {\bibfnamefont {I.}~\bibnamefont {Martinez-Soler}},\ and\
  \bibinfo {author} {\bibfnamefont {T.}~\bibnamefont {Schwetz}},\ }\bibfield
  {title} {\bibinfo {title} {{Updated Global Analysis of Neutrino Oscillations
  in the Presence of eV-Scale Sterile Neutrinos}},\ }\href
  {https://doi.org/10.1007/JHEP08(2018)010} {\bibfield  {journal} {\bibinfo
  {journal} {JHEP}\ }\textbf {\bibinfo {volume} {08}},\ \bibinfo {pages}
  {010}},\ \Eprint {https://arxiv.org/abs/1803.10661} {arXiv:1803.10661
  [hep-ph]} \BibitemShut {NoStop}%
\bibitem [{\citenamefont {Aartsen}\ \emph
  {et~al.}(2020{\natexlab{a}})\citenamefont {Aartsen} \emph
  {et~al.}}]{MEOWSPRD}%
  \BibitemOpen
  \bibfield  {author} {\bibinfo {author} {\bibfnamefont {M.~G.}\ \bibnamefont
  {Aartsen}} \emph {et~al.} (\bibinfo {collaboration} {IceCube}),\ }\bibfield
  {title} {\bibinfo {title} {{Searching for eV-scale sterile neutrinos with
  eight years of atmospheric neutrinos at the IceCube neutrino telescope}},\
  }\href@noop {} {\  (\bibinfo {year} {2020}{\natexlab{a}})}\BibitemShut
  {NoStop}%
\bibitem [{\citenamefont {Aartsen}\ \emph
  {et~al.}(2017{\natexlab{b}})\citenamefont {Aartsen} \emph
  {et~al.}}]{Aartsen:2016nxy}%
  \BibitemOpen
  \bibfield  {author} {\bibinfo {author} {\bibfnamefont {M.~G.}\ \bibnamefont
  {Aartsen}} \emph {et~al.} (\bibinfo {collaboration} {IceCube}),\ }\bibfield
  {title} {\bibinfo {title} {{The IceCube Neutrino Observatory: Instrumentation
  and Online Systems}},\ }\href
  {https://doi.org/10.1088/1748-0221/12/03/P03012} {\bibfield  {journal}
  {\bibinfo  {journal} {JINST}\ }\textbf {\bibinfo {volume} {12}}\bibfield
  {number} {\bibinfo  {number} { (03)},\ \bibinfo {pages} {P03012}},\ }\Eprint
  {https://arxiv.org/abs/1612.05093} {arXiv:1612.05093 [astro-ph.IM]}
  \BibitemShut {NoStop}%
\bibitem [{\citenamefont {Abbasi}\ \emph {et~al.}(2009)\citenamefont {Abbasi}
  \emph {et~al.}}]{Abbasi:2008aa}%
  \BibitemOpen
  \bibfield  {author} {\bibinfo {author} {\bibfnamefont {R.}~\bibnamefont
  {Abbasi}} \emph {et~al.} (\bibinfo {collaboration} {IceCube}),\ }\bibfield
  {title} {\bibinfo {title} {{The IceCube Data Acquisition System: Signal
  Capture, Digitization, and Timestamping}},\ }\href
  {https://doi.org/10.1016/j.nima.2009.01.001} {\bibfield  {journal} {\bibinfo
  {journal} {Nucl. Instrum. Meth.}\ }\textbf {\bibinfo {volume} {A601}},\
  \bibinfo {pages} {294} (\bibinfo {year} {2009})},\ \Eprint
  {https://arxiv.org/abs/0810.4930} {arXiv:0810.4930 [physics.ins-det]}
  \BibitemShut {NoStop}%
\bibitem [{\citenamefont {Abbasi}\ \emph {et~al.}(2012)\citenamefont {Abbasi}
  \emph {et~al.}}]{Collaboration:2011ym}%
  \BibitemOpen
  \bibfield  {author} {\bibinfo {author} {\bibfnamefont {R.}~\bibnamefont
  {Abbasi}} \emph {et~al.} (\bibinfo {collaboration} {IceCube}),\ }\bibfield
  {title} {\bibinfo {title} {{The Design and Performance of IceCube
  DeepCore}},\ }\href {https://doi.org/10.1016/j.astropartphys.2012.01.004}
  {\bibfield  {journal} {\bibinfo  {journal} {Astropart. Phys.}\ }\textbf
  {\bibinfo {volume} {35}},\ \bibinfo {pages} {615} (\bibinfo {year} {2012})},\
  \Eprint {https://arxiv.org/abs/1109.6096} {arXiv:1109.6096 [astro-ph.IM]}
  \BibitemShut {NoStop}%
\bibitem [{\citenamefont {Bugaev}\ \emph {et~al.}(1998)\citenamefont {Bugaev},
  \citenamefont {Misaki}, \citenamefont {Naumov}, \citenamefont {Sinegovskaya},
  \citenamefont {Sinegovsky},\ and\ \citenamefont {Takahashi}}]{Bugaev:1998bi}%
  \BibitemOpen
  \bibfield  {author} {\bibinfo {author} {\bibfnamefont {E.~V.}\ \bibnamefont
  {Bugaev}}, \bibinfo {author} {\bibfnamefont {A.}~\bibnamefont {Misaki}},
  \bibinfo {author} {\bibfnamefont {V.~A.}\ \bibnamefont {Naumov}}, \bibinfo
  {author} {\bibfnamefont {T.~S.}\ \bibnamefont {Sinegovskaya}}, \bibinfo
  {author} {\bibfnamefont {S.~I.}\ \bibnamefont {Sinegovsky}},\ and\ \bibinfo
  {author} {\bibfnamefont {N.}~\bibnamefont {Takahashi}},\ }\bibfield  {title}
  {\bibinfo {title} {{Atmospheric muon flux at sea level, underground and
  underwater}},\ }\href {https://doi.org/10.1103/PhysRevD.58.054001} {\bibfield
   {journal} {\bibinfo  {journal} {Phys. Rev.}\ }\textbf {\bibinfo {volume}
  {D58}},\ \bibinfo {pages} {054001} (\bibinfo {year} {1998})},\ \Eprint
  {https://arxiv.org/abs/hep-ph/9803488} {arXiv:hep-ph/9803488 [hep-ph]}
  \BibitemShut {NoStop}%
\bibitem [{\citenamefont {Gandhi}\ \emph {et~al.}(1996)\citenamefont {Gandhi},
  \citenamefont {Quigg}, \citenamefont {Reno},\ and\ \citenamefont
  {Sarcevic}}]{Gandhi:1995tf}%
  \BibitemOpen
  \bibfield  {author} {\bibinfo {author} {\bibfnamefont {R.}~\bibnamefont
  {Gandhi}}, \bibinfo {author} {\bibfnamefont {C.}~\bibnamefont {Quigg}},
  \bibinfo {author} {\bibfnamefont {M.~H.}\ \bibnamefont {Reno}},\ and\
  \bibinfo {author} {\bibfnamefont {I.}~\bibnamefont {Sarcevic}},\ }\bibfield
  {title} {\bibinfo {title} {{Ultrahigh-energy neutrino interactions}},\ }\href
  {https://doi.org/10.1016/0927-6505(96)00008-4} {\bibfield  {journal}
  {\bibinfo  {journal} {Astropart. Phys.}\ }\textbf {\bibinfo {volume} {5}},\
  \bibinfo {pages} {81} (\bibinfo {year} {1996})},\ \Eprint
  {https://arxiv.org/abs/hep-ph/9512364} {arXiv:hep-ph/9512364 [hep-ph]}
  \BibitemShut {NoStop}%
\bibitem [{\citenamefont {Bramall}\ \emph {et~al.}(2005)\citenamefont
  {Bramall}, \citenamefont {Bay}, \citenamefont {Woschnagg}, \citenamefont
  {Rohde},\ and\ \citenamefont {Price}}]{GRL:GRL20535}%
  \BibitemOpen
  \bibfield  {author} {\bibinfo {author} {\bibfnamefont {N.~E.}\ \bibnamefont
  {Bramall}}, \bibinfo {author} {\bibfnamefont {R.~C.}\ \bibnamefont {Bay}},
  \bibinfo {author} {\bibfnamefont {K.}~\bibnamefont {Woschnagg}}, \bibinfo
  {author} {\bibfnamefont {R.~A.}\ \bibnamefont {Rohde}},\ and\ \bibinfo
  {author} {\bibfnamefont {P.~B.}\ \bibnamefont {Price}},\ }\bibfield  {title}
  {\bibinfo {title} {A deep high-resolution optical log of dust, ash, and
  stratigraphy in south pole glacial ice},\ }\bibfield  {journal} {\bibinfo
  {journal} {Geophysical Research Letters}\ }\textbf {\bibinfo {volume} {32}},\
  \href {https://doi.org/10.1029/2005GL024236} {10.1029/2005GL024236} (\bibinfo
  {year} {2005})\BibitemShut {NoStop}%
\bibitem [{\citenamefont {Askebjer}\ \emph {et~al.}(1997)\citenamefont
  {Askebjer} \emph {et~al.}}]{Askebjer:1997ep}%
  \BibitemOpen
  \bibfield  {author} {\bibinfo {author} {\bibfnamefont {P.}~\bibnamefont
  {Askebjer}} \emph {et~al.},\ }\bibfield  {title} {\bibinfo {title} {{Optical
  properties of deep ice at the South Pole: Absorption}},\ }\href
  {https://doi.org/10.1364/AO.36.004168} {\bibfield  {journal} {\bibinfo
  {journal} {Appl. Opt.}\ }\textbf {\bibinfo {volume} {36}},\ \bibinfo {pages}
  {4168} (\bibinfo {year} {1997})},\ \Eprint
  {https://arxiv.org/abs/physics/9701025} {arXiv:physics/9701025 [physics]}
  \BibitemShut {NoStop}%
\bibitem [{\citenamefont {Aartsen}\ \emph
  {et~al.}(2013{\natexlab{a}})\citenamefont {Aartsen} \emph
  {et~al.}}]{Aartsen:2013rt}%
  \BibitemOpen
  \bibfield  {author} {\bibinfo {author} {\bibfnamefont {M.~G.}\ \bibnamefont
  {Aartsen}} \emph {et~al.} (\bibinfo {collaboration} {IceCube}),\ }\bibfield
  {title} {\bibinfo {title} {{Measurement of South Pole ice transparency with
  the IceCube LED calibration system}},\ }\href
  {https://doi.org/10.1016/j.nima.2013.01.054} {\bibfield  {journal} {\bibinfo
  {journal} {Nucl. Instrum. Meth.}\ }\textbf {\bibinfo {volume} {A711}},\
  \bibinfo {pages} {73} (\bibinfo {year} {2013}{\natexlab{a}})},\ \Eprint
  {https://arxiv.org/abs/1301.5361} {arXiv:1301.5361 [astro-ph.IM]}
  \BibitemShut {NoStop}%
\bibitem [{\citenamefont {Lipari}(1993)}]{Lipari:1993hd}%
  \BibitemOpen
  \bibfield  {author} {\bibinfo {author} {\bibfnamefont {P.}~\bibnamefont
  {Lipari}},\ }\bibfield  {title} {\bibinfo {title} {{Lepton spectra in the
  earth's atmosphere}},\ }\href {https://doi.org/10.1016/0927-6505(93)90022-6}
  {\bibfield  {journal} {\bibinfo  {journal} {Astropart. Phys.}\ }\textbf
  {\bibinfo {volume} {1}},\ \bibinfo {pages} {195} (\bibinfo {year}
  {1993})}\BibitemShut {NoStop}%
\bibitem [{\citenamefont {Koehne}\ \emph {et~al.}(2013)\citenamefont {Koehne},
  \citenamefont {Frantzen}, \citenamefont {Schmitz}, \citenamefont {Fuchs},
  \citenamefont {Rhode}, \citenamefont {Chirkin},\ and\ \citenamefont
  {Becker~Tjus}}]{Koehne:2013gpa}%
  \BibitemOpen
  \bibfield  {author} {\bibinfo {author} {\bibfnamefont {J.~H.}\ \bibnamefont
  {Koehne}}, \bibinfo {author} {\bibfnamefont {K.}~\bibnamefont {Frantzen}},
  \bibinfo {author} {\bibfnamefont {M.}~\bibnamefont {Schmitz}}, \bibinfo
  {author} {\bibfnamefont {T.}~\bibnamefont {Fuchs}}, \bibinfo {author}
  {\bibfnamefont {W.}~\bibnamefont {Rhode}}, \bibinfo {author} {\bibfnamefont
  {D.}~\bibnamefont {Chirkin}},\ and\ \bibinfo {author} {\bibfnamefont
  {J.}~\bibnamefont {Becker~Tjus}},\ }\bibfield  {title} {\bibinfo {title}
  {{PROPOSAL: A tool for propagation of charged leptons}},\ }\href
  {https://doi.org/10.1016/j.cpc.2013.04.001} {\bibfield  {journal} {\bibinfo
  {journal} {Comput. Phys. Commun.}\ }\textbf {\bibinfo {volume} {184}},\
  \bibinfo {pages} {2070} (\bibinfo {year} {2013})}\BibitemShut {NoStop}%
\bibitem [{\citenamefont {Halzen}(2006)}]{Halzen:2006mq}%
  \BibitemOpen
  \bibfield  {author} {\bibinfo {author} {\bibfnamefont {F.}~\bibnamefont
  {Halzen}},\ }\bibfield  {title} {\bibinfo {title} {{Astroparticle physics
  with high energy neutrinos: from AMANDA to IceCube}},\ }\href
  {https://doi.org/10.1140/epjc/s2006-02536-4} {\bibfield  {journal} {\bibinfo
  {journal} {Eur. Phys. J.}\ }\textbf {\bibinfo {volume} {C46}},\ \bibinfo
  {pages} {669} (\bibinfo {year} {2006})},\ \Eprint
  {https://arxiv.org/abs/astro-ph/0602132} {arXiv:astro-ph/0602132 [astro-ph]}
  \BibitemShut {NoStop}%
\bibitem [{\citenamefont {Aartsen}\ \emph {et~al.}(2015)\citenamefont {Aartsen}
  \emph {et~al.}}]{Aartsen:2015rwa}%
  \BibitemOpen
  \bibfield  {author} {\bibinfo {author} {\bibfnamefont {M.}~\bibnamefont
  {Aartsen}} \emph {et~al.} (\bibinfo {collaboration} {IceCube}),\ }\bibfield
  {title} {\bibinfo {title} {{Evidence for Astrophysical Muon Neutrinos from
  the Northern Sky with IceCube}},\ }\href
  {https://doi.org/10.1103/PhysRevLett.115.081102} {\bibfield  {journal}
  {\bibinfo  {journal} {Phys. Rev. Lett.}\ }\textbf {\bibinfo {volume} {115}},\
  \bibinfo {pages} {081102} (\bibinfo {year} {2015})},\ \Eprint
  {https://arxiv.org/abs/1507.04005} {arXiv:1507.04005 [astro-ph.HE]}
  \BibitemShut {NoStop}%
\bibitem [{\citenamefont {Ahrens}\ \emph {et~al.}(2004)\citenamefont {Ahrens}
  \emph {et~al.}}]{Ahrens:2003fg}%
  \BibitemOpen
  \bibfield  {author} {\bibinfo {author} {\bibfnamefont {J.}~\bibnamefont
  {Ahrens}} \emph {et~al.} (\bibinfo {collaboration} {AMANDA}),\ }\bibfield
  {title} {\bibinfo {title} {{Muon track reconstruction and data selection
  techniques in AMANDA}},\ }\href {https://doi.org/10.1016/j.nima.2004.01.065}
  {\bibfield  {journal} {\bibinfo  {journal} {Nucl. Instrum. Meth.}\ }\textbf
  {\bibinfo {volume} {A524}},\ \bibinfo {pages} {169} (\bibinfo {year}
  {2004})},\ \Eprint {https://arxiv.org/abs/astro-ph/0407044}
  {arXiv:astro-ph/0407044 [astro-ph]} \BibitemShut {NoStop}%
\bibitem [{\citenamefont {Aartsen}\ \emph
  {et~al.}(2014{\natexlab{a}})\citenamefont {Aartsen} \emph
  {et~al.}}]{Aartsen:2013vja}%
  \BibitemOpen
  \bibfield  {author} {\bibinfo {author} {\bibfnamefont {M.~G.}\ \bibnamefont
  {Aartsen}} \emph {et~al.} (\bibinfo {collaboration} {IceCube}),\ }\bibfield
  {title} {\bibinfo {title} {{Energy Reconstruction Methods in the IceCube
  Neutrino Telescope}},\ }\href {https://doi.org/10.1088/1748-0221/9/03/P03009}
  {\bibfield  {journal} {\bibinfo  {journal} {JINST}\ }\textbf {\bibinfo
  {volume} {9}},\ \bibinfo {pages} {P03009}},\ \Eprint
  {https://arxiv.org/abs/1311.4767} {arXiv:1311.4767 [physics.ins-det]}
  \BibitemShut {NoStop}%
\bibitem [{\citenamefont {Heck}\ \emph {et~al.}(1998)\citenamefont {Heck},
  \citenamefont {Knapp}, \citenamefont {Capdevielle}, \citenamefont {Schatz},\
  and\ \citenamefont {Thouw}}]{Heck:1998vt}%
  \BibitemOpen
  \bibfield  {author} {\bibinfo {author} {\bibfnamefont {D.}~\bibnamefont
  {Heck}}, \bibinfo {author} {\bibfnamefont {J.}~\bibnamefont {Knapp}},
  \bibinfo {author} {\bibfnamefont {J.~N.}\ \bibnamefont {Capdevielle}},
  \bibinfo {author} {\bibfnamefont {G.}~\bibnamefont {Schatz}},\ and\ \bibinfo
  {author} {\bibfnamefont {T.}~\bibnamefont {Thouw}},\ }\bibfield  {title}
  {\bibinfo {title} {{CORSIKA: A Monte Carlo code to simulate extensive air
  showers}},\ }\href@noop {} {\  (\bibinfo {year} {1998})}\BibitemShut
  {NoStop}%
\bibitem [{\citenamefont {Fedynitch}\ \emph {et~al.}(2015)\citenamefont
  {Fedynitch}, \citenamefont {Engel}, \citenamefont {Gaisser}, \citenamefont
  {Riehn},\ and\ \citenamefont {Stanev}}]{Fedynitch:2015zma}%
  \BibitemOpen
  \bibfield  {author} {\bibinfo {author} {\bibfnamefont {A.}~\bibnamefont
  {Fedynitch}}, \bibinfo {author} {\bibfnamefont {R.}~\bibnamefont {Engel}},
  \bibinfo {author} {\bibfnamefont {T.~K.}\ \bibnamefont {Gaisser}}, \bibinfo
  {author} {\bibfnamefont {F.}~\bibnamefont {Riehn}},\ and\ \bibinfo {author}
  {\bibfnamefont {T.}~\bibnamefont {Stanev}},\ }\bibfield  {title} {\bibinfo
  {title} {{Calculation of conventional and prompt lepton fluxes at very high
  energy}},\ }\bibfield  {booktitle} {\emph {\bibinfo {booktitle}
  {{Proceedings, 18th International Symposium on Very High Energy Cosmic Ray
  Interactions (ISVHECRI 2014): Geneva, Switzerland, August 18-22, 2014}}},\
  }\href {https://doi.org/10.1051/epjconf/20159908001} {\bibfield  {journal}
  {\bibinfo  {journal} {EPJ Web Conf.}\ }\textbf {\bibinfo {volume} {99}},\
  \bibinfo {pages} {08001} (\bibinfo {year} {2015})},\ \Eprint
  {https://arxiv.org/abs/1503.00544} {arXiv:1503.00544 [hep-ph]} \BibitemShut
  {NoStop}%
\bibitem [{\citenamefont {Fedynitch}\ \emph {et~al.}(2012)\citenamefont
  {Fedynitch}, \citenamefont {Becker~Tjus},\ and\ \citenamefont
  {Desiati}}]{Fedynitch:2012fs}%
  \BibitemOpen
  \bibfield  {author} {\bibinfo {author} {\bibfnamefont {A.}~\bibnamefont
  {Fedynitch}}, \bibinfo {author} {\bibfnamefont {J.}~\bibnamefont
  {Becker~Tjus}},\ and\ \bibinfo {author} {\bibfnamefont {P.}~\bibnamefont
  {Desiati}},\ }\bibfield  {title} {\bibinfo {title} {{Influence of hadronic
  interaction models and the cosmic ray spectrum on the high energy atmospheric
  muon and neutrino flux}},\ }\href
  {https://doi.org/10.1103/PhysRevD.86.114024} {\bibfield  {journal} {\bibinfo
  {journal} {Phys. Rev.}\ }\textbf {\bibinfo {volume} {D86}},\ \bibinfo {pages}
  {114024} (\bibinfo {year} {2012})},\ \Eprint
  {https://arxiv.org/abs/1206.6710} {arXiv:1206.6710 [astro-ph.HE]}
  \BibitemShut {NoStop}%
\bibitem [{\citenamefont {Riehn}\ \emph {et~al.}(2018)\citenamefont {Riehn},
  \citenamefont {Dembinski}, \citenamefont {Engel}, \citenamefont {Fedynitch},
  \citenamefont {Gaisser},\ and\ \citenamefont {Stanev}}]{Riehn:2017mfm}%
  \BibitemOpen
  \bibfield  {author} {\bibinfo {author} {\bibfnamefont {F.}~\bibnamefont
  {Riehn}}, \bibinfo {author} {\bibfnamefont {H.~P.}\ \bibnamefont
  {Dembinski}}, \bibinfo {author} {\bibfnamefont {R.}~\bibnamefont {Engel}},
  \bibinfo {author} {\bibfnamefont {A.}~\bibnamefont {Fedynitch}}, \bibinfo
  {author} {\bibfnamefont {T.~K.}\ \bibnamefont {Gaisser}},\ and\ \bibinfo
  {author} {\bibfnamefont {T.}~\bibnamefont {Stanev}},\ }\bibfield  {title}
  {\bibinfo {title} {{The hadronic interaction model SIBYLL 2.3c and Feynman
  scaling}},\ }\bibfield  {booktitle} {\emph {\bibinfo {booktitle} {{The
  Fluorescence detector Array of Single-pixel Telescopes: Contributions to the
  35th International Cosmic Ray Conference (ICRC 2017)}}},\ }\href
  {https://doi.org/10.22323/1.301.0301} {\bibfield  {journal} {\bibinfo
  {journal} {PoS}\ }\textbf {\bibinfo {volume} {ICRC2017}},\ \bibinfo {pages}
  {301} (\bibinfo {year} {2018})},\ \bibinfo {note} {[35,301(2017)]},\ \Eprint
  {https://arxiv.org/abs/1709.07227} {arXiv:1709.07227 [hep-ph]} \BibitemShut
  {NoStop}%
\bibitem [{\citenamefont {Gaisser}(2012)}]{Gaisser:2011cc}%
  \BibitemOpen
  \bibfield  {author} {\bibinfo {author} {\bibfnamefont {T.~K.}\ \bibnamefont
  {Gaisser}},\ }\bibfield  {title} {\bibinfo {title} {{Spectrum of cosmic-ray
  nucleons, kaon production, and the atmospheric muon charge ratio}},\ }\href
  {https://doi.org/10.1016/j.astropartphys.2012.02.010} {\bibfield  {journal}
  {\bibinfo  {journal} {Astropart. Phys.}\ }\textbf {\bibinfo {volume} {35}},\
  \bibinfo {pages} {801} (\bibinfo {year} {2012})},\ \Eprint
  {https://arxiv.org/abs/1111.6675} {arXiv:1111.6675 [astro-ph.HE]}
  \BibitemShut {NoStop}%
\bibitem [{\citenamefont {Hillas}(2006)}]{Hillas:2006ms}%
  \BibitemOpen
  \bibfield  {author} {\bibinfo {author} {\bibfnamefont {A.~M.}\ \bibnamefont
  {Hillas}},\ }\bibfield  {title} {\bibinfo {title} {{Cosmic Rays: Recent
  Progress and some Current Questions}},\ }in\ \href@noop {} {\emph {\bibinfo
  {booktitle} {{Conference on Cosmology, Galaxy Formation and Astro-Particle
  Physics on the Pathway to the SKA Oxford, England, April 10-12, 2006}}}}\
  (\bibinfo {year} {2006})\ \Eprint {https://arxiv.org/abs/astro-ph/0607109}
  {arXiv:astro-ph/0607109 [astro-ph]} \BibitemShut {NoStop}%
\bibitem [{\citenamefont {{Jet Propulsion Laboratory }}(2014)}]{AIRS}%
  \BibitemOpen
  \bibfield  {author} {\bibinfo {author} {\bibnamefont {{Jet Propulsion
  Laboratory }}},\ }\bibfield  {title} {\bibinfo {title} {{AIRS/AMSU/HSB
  Version 6 Level 3 Product User Guide}},\ }\href@noop {} {\bibfield  {journal}
  {\bibinfo  {journal} {Version 1.2}\ } (\bibinfo {year} {November
  2014})}\BibitemShut {NoStop}%
\bibitem [{\citenamefont {Bhattacharya}\ \emph {et~al.}(2016)\citenamefont
  {Bhattacharya}, \citenamefont {Enberg}, \citenamefont {Jeong}, \citenamefont
  {Kim}, \citenamefont {Reno}, \citenamefont {Sarcevic},\ and\ \citenamefont
  {Stasto}}]{Bhattacharya:2016jce}%
  \BibitemOpen
  \bibfield  {author} {\bibinfo {author} {\bibfnamefont {A.}~\bibnamefont
  {Bhattacharya}}, \bibinfo {author} {\bibfnamefont {R.}~\bibnamefont
  {Enberg}}, \bibinfo {author} {\bibfnamefont {Y.~S.}\ \bibnamefont {Jeong}},
  \bibinfo {author} {\bibfnamefont {C.~S.}\ \bibnamefont {Kim}}, \bibinfo
  {author} {\bibfnamefont {M.~H.}\ \bibnamefont {Reno}}, \bibinfo {author}
  {\bibfnamefont {I.}~\bibnamefont {Sarcevic}},\ and\ \bibinfo {author}
  {\bibfnamefont {A.}~\bibnamefont {Stasto}},\ }\bibfield  {title} {\bibinfo
  {title} {{Prompt atmospheric neutrino fluxes: perturbative QCD models and
  nuclear effects}},\ }\href {https://doi.org/10.1007/JHEP11(2016)167}
  {\bibfield  {journal} {\bibinfo  {journal} {JHEP}\ }\textbf {\bibinfo
  {volume} {11}},\ \bibinfo {pages} {167}},\ \Eprint
  {https://arxiv.org/abs/1607.00193} {arXiv:1607.00193 [hep-ph]} \BibitemShut
  {NoStop}%
\bibitem [{\citenamefont {Palladino}\ and\ \citenamefont
  {Vissani}(2015)}]{Palladino:2015vna}%
  \BibitemOpen
  \bibfield  {author} {\bibinfo {author} {\bibfnamefont {A.}~\bibnamefont
  {Palladino}}\ and\ \bibinfo {author} {\bibfnamefont {F.}~\bibnamefont
  {Vissani}},\ }\bibfield  {title} {\bibinfo {title} {{The natural
  parameterization of cosmic neutrino oscillations}},\ }\href
  {https://doi.org/10.1140/epjc/s10052-015-3664-6} {\bibfield  {journal}
  {\bibinfo  {journal} {Eur. Phys. J.}\ }\textbf {\bibinfo {volume} {C75}},\
  \bibinfo {pages} {433} (\bibinfo {year} {2015})},\ \Eprint
  {https://arxiv.org/abs/1504.05238} {arXiv:1504.05238 [hep-ph]} \BibitemShut
  {NoStop}%
\bibitem [{\citenamefont {Argüelles}\ \emph {et~al.}(2015)\citenamefont
  {Argüelles}, \citenamefont {Katori},\ and\ \citenamefont
  {Salvado}}]{Arguelles:2015dca}%
  \BibitemOpen
  \bibfield  {author} {\bibinfo {author} {\bibfnamefont {C.~A.}\ \bibnamefont
  {Argüelles}}, \bibinfo {author} {\bibfnamefont {T.}~\bibnamefont {Katori}},\
  and\ \bibinfo {author} {\bibfnamefont {J.}~\bibnamefont {Salvado}},\
  }\bibfield  {title} {\bibinfo {title} {{New Physics in Astrophysical Neutrino
  Flavor}},\ }\href {https://doi.org/10.1103/PhysRevLett.115.161303} {\bibfield
   {journal} {\bibinfo  {journal} {Phys. Rev. Lett.}\ }\textbf {\bibinfo
  {volume} {115}},\ \bibinfo {pages} {161303} (\bibinfo {year} {2015})},\
  \Eprint {https://arxiv.org/abs/1506.02043} {arXiv:1506.02043 [hep-ph]}
  \BibitemShut {NoStop}%
\bibitem [{\citenamefont {Bustamante}\ \emph {et~al.}(2015)\citenamefont
  {Bustamante}, \citenamefont {Beacom},\ and\ \citenamefont
  {Winter}}]{Bustamante:2015waa}%
  \BibitemOpen
  \bibfield  {author} {\bibinfo {author} {\bibfnamefont {M.}~\bibnamefont
  {Bustamante}}, \bibinfo {author} {\bibfnamefont {J.~F.}\ \bibnamefont
  {Beacom}},\ and\ \bibinfo {author} {\bibfnamefont {W.}~\bibnamefont
  {Winter}},\ }\bibfield  {title} {\bibinfo {title} {{Theoretically palatable
  flavor combinations of astrophysical neutrinos}},\ }\href
  {https://doi.org/10.1103/PhysRevLett.115.161302} {\bibfield  {journal}
  {\bibinfo  {journal} {Phys. Rev. Lett.}\ }\textbf {\bibinfo {volume} {115}},\
  \bibinfo {pages} {161302} (\bibinfo {year} {2015})},\ \Eprint
  {https://arxiv.org/abs/1506.02645} {arXiv:1506.02645 [astro-ph.HE]}
  \BibitemShut {NoStop}%
\bibitem [{\citenamefont {Schneider}(2019)}]{Schneider:2019ayi}%
  \BibitemOpen
  \bibfield  {author} {\bibinfo {author} {\bibfnamefont {A.}~\bibnamefont
  {Schneider}},\ }\bibfield  {title} {\bibinfo {title} {{Characterization of
  the Astrophysical Diffuse Neutrino Flux with IceCube High-Energy Starting
  Events}},\ }in\ \href@noop {} {\emph {\bibinfo {booktitle} {{36th
  International Cosmic Ray Conference (ICRC 2019) Madison, Wisconsin, USA, July
  24-August 1, 2019}}}}\ (\bibinfo {year} {2019})\ \Eprint
  {https://arxiv.org/abs/1907.11266} {arXiv:1907.11266 [astro-ph.HE]}
  \BibitemShut {NoStop}%
\bibitem [{\citenamefont {Argüelles~Delgado}\ \emph
  {et~al.}(2015)\citenamefont {Argüelles~Delgado}, \citenamefont {Salvado},\
  and\ \citenamefont {Weaver}}]{Delgado:2014kpa}%
  \BibitemOpen
  \bibfield  {author} {\bibinfo {author} {\bibfnamefont {C.~A.}\ \bibnamefont
  {Argüelles~Delgado}}, \bibinfo {author} {\bibfnamefont {J.}~\bibnamefont
  {Salvado}},\ and\ \bibinfo {author} {\bibfnamefont {C.~N.}\ \bibnamefont
  {Weaver}},\ }\bibfield  {title} {\bibinfo {title} {{A Simple Quantum
  Integro-Differential Solver (SQuIDS)}},\ }\href
  {https://doi.org/10.1016/j.cpc.2015.06.022} {\bibfield  {journal} {\bibinfo
  {journal} {Comput. Phys. Commun.}\ }\textbf {\bibinfo {volume} {196}},\
  \bibinfo {pages} {569} (\bibinfo {year} {2015})},\ \Eprint
  {https://arxiv.org/abs/1412.3832} {arXiv:1412.3832 [hep-ph]} \BibitemShut
  {NoStop}%
\bibitem [{\citenamefont {Arg\"uelles}\ \emph {et~al.}(2015)\citenamefont
  {Arg\"uelles}, \citenamefont {Salvado},\ and\ \citenamefont
  {Weaver}}]{nusquids}%
  \BibitemOpen
  \bibfield  {author} {\bibinfo {author} {\bibfnamefont {C.~A.}\ \bibnamefont
  {Arg\"uelles}}, \bibinfo {author} {\bibfnamefont {J.}~\bibnamefont
  {Salvado}},\ and\ \bibinfo {author} {\bibfnamefont {C.~N.}\ \bibnamefont
  {Weaver}},\ }\href@noop {} {\bibinfo {title} {{nuSQuIDS}}},\ \bibinfo
  {howpublished} {\url{https://github.com/Arguelles/nuSQuIDS}} (\bibinfo {year}
  {2015})\BibitemShut {NoStop}%
\bibitem [{\citenamefont {Gonzalez-Garcia}\ \emph {et~al.}(2005)\citenamefont
  {Gonzalez-Garcia}, \citenamefont {Halzen},\ and\ \citenamefont
  {Maltoni}}]{GonzalezGarcia:2005xw}%
  \BibitemOpen
  \bibfield  {author} {\bibinfo {author} {\bibfnamefont {M.~C.}\ \bibnamefont
  {Gonzalez-Garcia}}, \bibinfo {author} {\bibfnamefont {F.}~\bibnamefont
  {Halzen}},\ and\ \bibinfo {author} {\bibfnamefont {M.}~\bibnamefont
  {Maltoni}},\ }\bibfield  {title} {\bibinfo {title} {{Physics reach of
  high-energy and high-statistics IceCube atmospheric neutrino data}},\ }\href
  {https://doi.org/10.1103/PhysRevD.71.093010} {\bibfield  {journal} {\bibinfo
  {journal} {Phys. Rev.}\ }\textbf {\bibinfo {volume} {D71}},\ \bibinfo {pages}
  {093010} (\bibinfo {year} {2005})},\ \Eprint
  {https://arxiv.org/abs/hep-ph/0502223} {arXiv:hep-ph/0502223 [hep-ph]}
  \BibitemShut {NoStop}%
\bibitem [{\citenamefont {Glashow}(1960)}]{Glashow:1960zz}%
  \BibitemOpen
  \bibfield  {author} {\bibinfo {author} {\bibfnamefont {S.~L.}\ \bibnamefont
  {Glashow}},\ }\bibfield  {title} {\bibinfo {title} {{Resonant Scattering of
  Antineutrinos}},\ }\href {https://doi.org/10.1103/PhysRev.118.316} {\bibfield
   {journal} {\bibinfo  {journal} {Phys. Rev.}\ }\textbf {\bibinfo {volume}
  {118}},\ \bibinfo {pages} {316} (\bibinfo {year} {1960})}\BibitemShut
  {NoStop}%
\bibitem [{\citenamefont {Halzen}\ and\ \citenamefont
  {Saltzberg}(1998)}]{Halzen:1998be}%
  \BibitemOpen
  \bibfield  {author} {\bibinfo {author} {\bibfnamefont {F.}~\bibnamefont
  {Halzen}}\ and\ \bibinfo {author} {\bibfnamefont {D.}~\bibnamefont
  {Saltzberg}},\ }\bibfield  {title} {\bibinfo {title} {{Tau-neutrino
  appearance with a 1000 megaparsec baseline}},\ }\href
  {https://doi.org/10.1103/PhysRevLett.81.4305} {\bibfield  {journal} {\bibinfo
   {journal} {Phys. Rev. Lett.}\ }\textbf {\bibinfo {volume} {81}},\ \bibinfo
  {pages} {4305} (\bibinfo {year} {1998})},\ \Eprint
  {https://arxiv.org/abs/hep-ph/9804354} {arXiv:hep-ph/9804354 [hep-ph]}
  \BibitemShut {NoStop}%
\bibitem [{\citenamefont {Cooper-Sarkar}\ \emph {et~al.}(2011)\citenamefont
  {Cooper-Sarkar}, \citenamefont {Mertsch},\ and\ \citenamefont
  {Sarkar}}]{CooperSarkar:2011pa}%
  \BibitemOpen
  \bibfield  {author} {\bibinfo {author} {\bibfnamefont {A.}~\bibnamefont
  {Cooper-Sarkar}}, \bibinfo {author} {\bibfnamefont {P.}~\bibnamefont
  {Mertsch}},\ and\ \bibinfo {author} {\bibfnamefont {S.}~\bibnamefont
  {Sarkar}},\ }\bibfield  {title} {\bibinfo {title} {{The high energy neutrino
  cross-section in the Standard Model and its uncertainty}},\ }\href
  {https://doi.org/10.1007/JHEP08(2011)042} {\bibfield  {journal} {\bibinfo
  {journal} {JHEP}\ }\textbf {\bibinfo {volume} {08}},\ \bibinfo {pages}
  {042}},\ \Eprint {https://arxiv.org/abs/1106.3723} {arXiv:1106.3723 [hep-ph]}
  \BibitemShut {NoStop}%
\bibitem [{\citenamefont {Dziewonski}\ and\ \citenamefont
  {Anderson}(1981)}]{Dziewonski:1981xy}%
  \BibitemOpen
  \bibfield  {author} {\bibinfo {author} {\bibfnamefont {A.~M.}\ \bibnamefont
  {Dziewonski}}\ and\ \bibinfo {author} {\bibfnamefont {D.~L.}\ \bibnamefont
  {Anderson}},\ }\bibfield  {title} {\bibinfo {title} {{Preliminary reference
  earth model}},\ }\href {https://doi.org/10.1016/0031-9201(81)90046-7}
  {\bibfield  {journal} {\bibinfo  {journal} {Phys. Earth Planet. Interiors}\
  }\textbf {\bibinfo {volume} {25}},\ \bibinfo {pages} {297} (\bibinfo {year}
  {1981})}\BibitemShut {NoStop}%
\bibitem [{\citenamefont {Argüelles}\ \emph {et~al.}(2019)\citenamefont
  {Argüelles}, \citenamefont {Schneider},\ and\ \citenamefont
  {Yuan}}]{Arguelles:2019izp}%
  \BibitemOpen
  \bibfield  {author} {\bibinfo {author} {\bibfnamefont {C.~A.}\ \bibnamefont
  {Argüelles}}, \bibinfo {author} {\bibfnamefont {A.}~\bibnamefont
  {Schneider}},\ and\ \bibinfo {author} {\bibfnamefont {T.}~\bibnamefont
  {Yuan}},\ }\bibfield  {title} {\bibinfo {title} {{A binned likelihood for
  stochastic models}},\ }\href {https://doi.org/10.1007/JHEP06(2019)030}
  {\bibfield  {journal} {\bibinfo  {journal} {JHEP}\ }\textbf {\bibinfo
  {volume} {06}},\ \bibinfo {pages} {030}},\ \Eprint
  {https://arxiv.org/abs/1901.04645} {arXiv:1901.04645 [physics.data-an]}
  \BibitemShut {NoStop}%
\bibitem [{\citenamefont {Wilks}(1938)}]{wilks1938}%
  \BibitemOpen
  \bibfield  {author} {\bibinfo {author} {\bibfnamefont {S.~S.}\ \bibnamefont
  {Wilks}},\ }\bibfield  {title} {\bibinfo {title} {The large-sample
  distribution of the likelihood ratio for testing composite hypotheses},\
  }\href {https://doi.org/10.1214/aoms/1177732360} {\bibfield  {journal}
  {\bibinfo  {journal} {Ann. Math. Statist.}\ }\textbf {\bibinfo {volume}
  {9}},\ \bibinfo {pages} {60} (\bibinfo {year} {1938})}\BibitemShut {NoStop}%
\bibitem [{\citenamefont {Feldman}\ and\ \citenamefont
  {Cousins}(1998)}]{Feldman:1997qc}%
  \BibitemOpen
  \bibfield  {author} {\bibinfo {author} {\bibfnamefont {G.~J.}\ \bibnamefont
  {Feldman}}\ and\ \bibinfo {author} {\bibfnamefont {R.~D.}\ \bibnamefont
  {Cousins}},\ }\bibfield  {title} {\bibinfo {title} {{A Unified approach to
  the classical statistical analysis of small signals}},\ }\href
  {https://doi.org/10.1103/PhysRevD.57.3873} {\bibfield  {journal} {\bibinfo
  {journal} {Phys. Rev.}\ }\textbf {\bibinfo {volume} {D57}},\ \bibinfo {pages}
  {3873} (\bibinfo {year} {1998})},\ \Eprint
  {https://arxiv.org/abs/physics/9711021} {arXiv:physics/9711021
  [physics.data-an]} \BibitemShut {NoStop}%
\bibitem [{\citenamefont {von Toussaint}(2011)}]{RevModPhys.83.943}%
  \BibitemOpen
  \bibfield  {author} {\bibinfo {author} {\bibfnamefont {U.}~\bibnamefont {von
  Toussaint}},\ }\bibfield  {title} {\bibinfo {title} {Bayesian inference in
  physics},\ }\href {https://doi.org/10.1103/RevModPhys.83.943} {\bibfield
  {journal} {\bibinfo  {journal} {Rev. Mod. Phys.}\ }\textbf {\bibinfo {volume}
  {83}},\ \bibinfo {pages} {943} (\bibinfo {year} {2011})}\BibitemShut
  {NoStop}%
\bibitem [{\citenamefont {Feroz}\ \emph {et~al.}(2009)\citenamefont {Feroz},
  \citenamefont {Hobson},\ and\ \citenamefont {Bridges}}]{Feroz:2008xx}%
  \BibitemOpen
  \bibfield  {author} {\bibinfo {author} {\bibfnamefont {F.}~\bibnamefont
  {Feroz}}, \bibinfo {author} {\bibfnamefont {M.~P.}\ \bibnamefont {Hobson}},\
  and\ \bibinfo {author} {\bibfnamefont {M.}~\bibnamefont {Bridges}},\
  }\bibfield  {title} {\bibinfo {title} {{MultiNest: an efficient and robust
  Bayesian inference tool for cosmology and particle physics}},\ }\href
  {https://doi.org/10.1111/j.1365-2966.2009.14548.x} {\bibfield  {journal}
  {\bibinfo  {journal} {Mon. Not. Roy. Astron. Soc.}\ }\textbf {\bibinfo
  {volume} {398}},\ \bibinfo {pages} {1601} (\bibinfo {year} {2009})},\ \Eprint
  {https://arxiv.org/abs/0809.3437} {arXiv:0809.3437 [astro-ph]} \BibitemShut
  {NoStop}%
\bibitem [{\citenamefont {Karelin}\ \emph {et~al.}(2011)\citenamefont
  {Karelin}, \citenamefont {Borisov}, \citenamefont {Galper},\ and\
  \citenamefont {Voronov}}]{Karelin:2011zz}%
  \BibitemOpen
  \bibfield  {author} {\bibinfo {author} {\bibfnamefont {A.~V.}\ \bibnamefont
  {Karelin}}, \bibinfo {author} {\bibfnamefont {S.~V.}\ \bibnamefont
  {Borisov}}, \bibinfo {author} {\bibfnamefont {A.~M.}\ \bibnamefont
  {Galper}},\ and\ \bibinfo {author} {\bibfnamefont {S.~A.}\ \bibnamefont
  {Voronov}},\ }\bibfield  {title} {\bibinfo {title} {{The proton and helium
  cosmic ray spectra from 50-GeV to 15-TeV}},\ }\href
  {https://doi.org/10.5194/astra-7-235-2011} {\bibfield  {journal} {\bibinfo
  {journal} {Astrophys. Space Sci. Trans.}\ }\textbf {\bibinfo {volume} {7}},\
  \bibinfo {pages} {235} (\bibinfo {year} {2011})}\BibitemShut {NoStop}%
\bibitem [{\citenamefont {Bartoli}\ \emph {et~al.}(2015)\citenamefont {Bartoli}
  \emph {et~al.}}]{Bartoli:2015fhw}%
  \BibitemOpen
  \bibfield  {author} {\bibinfo {author} {\bibfnamefont {B.}~\bibnamefont
  {Bartoli}} \emph {et~al.} (\bibinfo {collaboration} {ARGO-YBJ}),\ }\bibfield
  {title} {\bibinfo {title} {{Cosmic ray proton plus helium energy spectrum
  measured by the ARGO-YBJ experiment in the energy range 3–300 TeV}},\
  }\href {https://doi.org/10.1103/PhysRevD.91.112017} {\bibfield  {journal}
  {\bibinfo  {journal} {Phys. Rev.}\ }\textbf {\bibinfo {volume} {D91}},\
  \bibinfo {pages} {112017} (\bibinfo {year} {2015})},\ \Eprint
  {https://arxiv.org/abs/1503.07136} {arXiv:1503.07136 [hep-ex]} \BibitemShut
  {NoStop}%
\bibitem [{\citenamefont {Yoon}\ \emph {et~al.}(2017)\citenamefont {Yoon} \emph
  {et~al.}}]{Yoon:2017qjx}%
  \BibitemOpen
  \bibfield  {author} {\bibinfo {author} {\bibfnamefont {Y.~S.}\ \bibnamefont
  {Yoon}} \emph {et~al.},\ }\bibfield  {title} {\bibinfo {title} {{Proton and
  Helium Spectra from the CREAM-III Flight}},\ }\href
  {https://doi.org/10.3847/1538-4357/aa68e4} {\bibfield  {journal} {\bibinfo
  {journal} {Astrophys. J.}\ }\textbf {\bibinfo {volume} {839}},\ \bibinfo
  {pages} {5} (\bibinfo {year} {2017})},\ \Eprint
  {https://arxiv.org/abs/1704.02512} {arXiv:1704.02512 [astro-ph.HE]}
  \BibitemShut {NoStop}%
\bibitem [{\citenamefont {Alfaro}\ \emph {et~al.}(2017)\citenamefont {Alfaro}
  \emph {et~al.}}]{Alfaro:2017cwx}%
  \BibitemOpen
  \bibfield  {author} {\bibinfo {author} {\bibfnamefont {R.}~\bibnamefont
  {Alfaro}} \emph {et~al.} (\bibinfo {collaboration} {HAWC}),\ }\bibfield
  {title} {\bibinfo {title} {{All-particle cosmic ray energy spectrum measured
  by the HAWC experiment from 10 to 500 TeV}},\ }\href
  {https://doi.org/10.1103/PhysRevD.96.122001} {\bibfield  {journal} {\bibinfo
  {journal} {Phys. Rev.}\ }\textbf {\bibinfo {volume} {D96}},\ \bibinfo {pages}
  {122001} (\bibinfo {year} {2017})},\ \Eprint
  {https://arxiv.org/abs/1710.00890} {arXiv:1710.00890 [astro-ph.HE]}
  \BibitemShut {NoStop}%
\bibitem [{\citenamefont {Barr}\ \emph {et~al.}(2006)\citenamefont {Barr},
  \citenamefont {Gaisser}, \citenamefont {Robbins},\ and\ \citenamefont
  {Stanev}}]{Barr:2006it}%
  \BibitemOpen
  \bibfield  {author} {\bibinfo {author} {\bibfnamefont {G.~D.}\ \bibnamefont
  {Barr}}, \bibinfo {author} {\bibfnamefont {T.~K.}\ \bibnamefont {Gaisser}},
  \bibinfo {author} {\bibfnamefont {S.}~\bibnamefont {Robbins}},\ and\ \bibinfo
  {author} {\bibfnamefont {T.}~\bibnamefont {Stanev}},\ }\bibfield  {title}
  {\bibinfo {title} {{Uncertainties in Atmospheric Neutrino Fluxes}},\ }\href
  {https://doi.org/10.1103/PhysRevD.74.094009} {\bibfield  {journal} {\bibinfo
  {journal} {Phys. Rev.}\ }\textbf {\bibinfo {volume} {D74}},\ \bibinfo {pages}
  {094009} (\bibinfo {year} {2006})},\ \Eprint
  {https://arxiv.org/abs/astro-ph/0611266} {arXiv:astro-ph/0611266 [astro-ph]}
  \BibitemShut {NoStop}%
\bibitem [{\citenamefont {Halzen}\ \emph {et~al.}(2012)\citenamefont {Halzen},
  \citenamefont {Igi}, \citenamefont {Ishida},\ and\ \citenamefont
  {Kim}}]{PhysRevD.85.074020}%
  \BibitemOpen
  \bibfield  {author} {\bibinfo {author} {\bibfnamefont {F.}~\bibnamefont
  {Halzen}}, \bibinfo {author} {\bibfnamefont {K.}~\bibnamefont {Igi}},
  \bibinfo {author} {\bibfnamefont {M.}~\bibnamefont {Ishida}},\ and\ \bibinfo
  {author} {\bibfnamefont {C.~S.}\ \bibnamefont {Kim}},\ }\bibfield  {title}
  {\bibinfo {title} {Total hadronic cross sections and
  ${\ensuremath{\pi}}^{\ensuremath{\mp}}{\ensuremath{\pi}}^{+}$ scattering},\
  }\href {https://doi.org/10.1103/PhysRevD.85.074020} {\bibfield  {journal}
  {\bibinfo  {journal} {Phys. Rev. D}\ }\textbf {\bibinfo {volume} {85}},\
  \bibinfo {pages} {074020} (\bibinfo {year} {2012})}\BibitemShut {NoStop}%
\bibitem [{\citenamefont {Aartsen}\ \emph
  {et~al.}(2013{\natexlab{b}})\citenamefont {Aartsen} \emph
  {et~al.}}]{Aartsen:2013jdh}%
  \BibitemOpen
  \bibfield  {author} {\bibinfo {author} {\bibfnamefont {M.~G.}\ \bibnamefont
  {Aartsen}} \emph {et~al.} (\bibinfo {collaboration} {IceCube}),\ }\bibfield
  {title} {\bibinfo {title} {{Evidence for High-Energy Extraterrestrial
  Neutrinos at the IceCube Detector}},\ }\href
  {https://doi.org/10.1126/science.1242856} {\bibfield  {journal} {\bibinfo
  {journal} {Science}\ }\textbf {\bibinfo {volume} {342}},\ \bibinfo {pages}
  {1242856} (\bibinfo {year} {2013}{\natexlab{b}})},\ \Eprint
  {https://arxiv.org/abs/1311.5238} {arXiv:1311.5238 [astro-ph.HE]}
  \BibitemShut {NoStop}%
\bibitem [{\citenamefont {Aartsen}\ \emph
  {et~al.}(2014{\natexlab{b}})\citenamefont {Aartsen} \emph
  {et~al.}}]{Aartsen:2014gkd}%
  \BibitemOpen
  \bibfield  {author} {\bibinfo {author} {\bibfnamefont {M.~G.}\ \bibnamefont
  {Aartsen}} \emph {et~al.} (\bibinfo {collaboration} {IceCube}),\ }\bibfield
  {title} {\bibinfo {title} {{Observation of High-Energy Astrophysical
  Neutrinos in Three Years of IceCube Data}},\ }\href
  {https://doi.org/10.1103/PhysRevLett.113.101101} {\bibfield  {journal}
  {\bibinfo  {journal} {Phys. Rev. Lett.}\ }\textbf {\bibinfo {volume} {113}},\
  \bibinfo {pages} {101101} (\bibinfo {year} {2014}{\natexlab{b}})},\ \Eprint
  {https://arxiv.org/abs/1405.5303} {arXiv:1405.5303 [astro-ph.HE]}
  \BibitemShut {NoStop}%
\bibitem [{\citenamefont {Aartsen}\ \emph
  {et~al.}(2016{\natexlab{b}})\citenamefont {Aartsen} \emph
  {et~al.}}]{Aartsen:2016xlq}%
  \BibitemOpen
  \bibfield  {author} {\bibinfo {author} {\bibfnamefont {M.~G.}\ \bibnamefont
  {Aartsen}} \emph {et~al.} (\bibinfo {collaboration} {IceCube}),\ }\bibfield
  {title} {\bibinfo {title} {{Observation and Characterization of a Cosmic Muon
  Neutrino Flux from the Northern Hemisphere using six years of IceCube
  data}},\ }\href {https://doi.org/10.3847/0004-637X/833/1/3} {\bibfield
  {journal} {\bibinfo  {journal} {Astrophys. J.}\ }\textbf {\bibinfo {volume}
  {833}},\ \bibinfo {pages} {3} (\bibinfo {year} {2016}{\natexlab{b}})},\
  \Eprint {https://arxiv.org/abs/1607.08006} {arXiv:1607.08006 [astro-ph.HE]}
  \BibitemShut {NoStop}%
\bibitem [{\citenamefont {Aartsen}\ \emph
  {et~al.}(2019{\natexlab{a}})\citenamefont {Aartsen} \emph
  {et~al.}}]{Aartsen:2018vez}%
  \BibitemOpen
  \bibfield  {author} {\bibinfo {author} {\bibfnamefont {M.~G.}\ \bibnamefont
  {Aartsen}} \emph {et~al.} (\bibinfo {collaboration} {IceCube}),\ }\bibfield
  {title} {\bibinfo {title} {{Measurements using the inelasticity distribution
  of multi-TeV neutrino interactions in IceCube}},\ }\href
  {https://doi.org/10.1103/PhysRevD.99.032004} {\bibfield  {journal} {\bibinfo
  {journal} {Phys. Rev.}\ }\textbf {\bibinfo {volume} {D99}},\ \bibinfo {pages}
  {032004} (\bibinfo {year} {2019}{\natexlab{a}})},\ \Eprint
  {https://arxiv.org/abs/1808.07629} {arXiv:1808.07629 [hep-ex]} \BibitemShut
  {NoStop}%
\bibitem [{\citenamefont {Aartsen}\ \emph
  {et~al.}(2020{\natexlab{b}})\citenamefont {Aartsen} \emph
  {et~al.}}]{Aartsen:2020aqd}%
  \BibitemOpen
  \bibfield  {author} {\bibinfo {author} {\bibfnamefont {M.}~\bibnamefont
  {Aartsen}} \emph {et~al.} (\bibinfo {collaboration} {IceCube}),\ }\bibfield
  {title} {\bibinfo {title} {{Characteristics of the diffuse astrophysical
  electron and tau neutrino flux with six years of IceCube high energy cascade
  data}},\ }\href@noop {} {\  (\bibinfo {year} {2020}{\natexlab{b}})},\ \Eprint
  {https://arxiv.org/abs/2001.09520} {arXiv:2001.09520 [astro-ph.HE]}
  \BibitemShut {NoStop}%
\bibitem [{\citenamefont {Aartsen}\ \emph
  {et~al.}(2019{\natexlab{b}})\citenamefont {Aartsen} \emph
  {et~al.}}]{Aartsen:2019jcj}%
  \BibitemOpen
  \bibfield  {author} {\bibinfo {author} {\bibfnamefont {M.~G.}\ \bibnamefont
  {Aartsen}} \emph {et~al.} (\bibinfo {collaboration} {IceCube}),\ }\bibfield
  {title} {\bibinfo {title} {{Efficient propagation of systematic uncertainties
  from calibration to analysis with the SnowStorm method in IceCube}},\ }\href
  {https://doi.org/10.1088/1475-7516/2019/10/048} {\bibfield  {journal}
  {\bibinfo  {journal} {JCAP}\ }\textbf {\bibinfo {volume} {1910}}\bibfield
  {number} {\bibinfo  {number} { (10)},\ \bibinfo {pages} {048}},\ }\Eprint
  {https://arxiv.org/abs/1909.01530} {arXiv:1909.01530 [hep-ex]} \BibitemShut
  {NoStop}%
\bibitem [{\citenamefont {Karle}(1994)}]{Karle:1994eua}%
  \BibitemOpen
  \bibfield  {author} {\bibinfo {author} {\bibfnamefont {A.}~\bibnamefont
  {Karle}},\ }\emph {\bibinfo {title} {{Entwicklung eines neuartigen
  atmosphärischen Tscherenkovdetektors und Messungen an hochenergetischer
  Kosmischer Strahlung zwischen 15 und 1000 TeV}}},\ \href@noop {} {Ph.D.
  thesis},\ \bibinfo  {school} {München Univ.} (\bibinfo {year}
  {1994})\BibitemShut {NoStop}%
\bibitem [{\citenamefont {Aartsen}\ \emph {et~al.}(2018)\citenamefont {Aartsen}
  \emph {et~al.}}]{Aartsen:2017nmd}%
  \BibitemOpen
  \bibfield  {author} {\bibinfo {author} {\bibfnamefont {M.~G.}\ \bibnamefont
  {Aartsen}} \emph {et~al.} (\bibinfo {collaboration} {IceCube}),\ }\bibfield
  {title} {\bibinfo {title} {{Measurement of Atmospheric Neutrino Oscillations
  at 6–56 GeV with IceCube DeepCore}},\ }\href
  {https://doi.org/10.1103/PhysRevLett.120.071801} {\bibfield  {journal}
  {\bibinfo  {journal} {Phys. Rev. Lett.}\ }\textbf {\bibinfo {volume} {120}},\
  \bibinfo {pages} {071801} (\bibinfo {year} {2018})},\ \Eprint
  {https://arxiv.org/abs/1707.07081} {arXiv:1707.07081 [hep-ex]} \BibitemShut
  {NoStop}%
\bibitem [{\citenamefont {Vincent}\ \emph {et~al.}(2017)\citenamefont
  {Vincent}, \citenamefont {Argüelles},\ and\ \citenamefont
  {Kheirandish}}]{Vincent:2017svp}%
  \BibitemOpen
  \bibfield  {author} {\bibinfo {author} {\bibfnamefont {A.~C.}\ \bibnamefont
  {Vincent}}, \bibinfo {author} {\bibfnamefont {C.~A.}\ \bibnamefont
  {Argüelles}},\ and\ \bibinfo {author} {\bibfnamefont {A.}~\bibnamefont
  {Kheirandish}},\ }\bibfield  {title} {\bibinfo {title} {{High-energy neutrino
  attenuation in the Earth and its associated uncertainties}},\ }\href
  {https://doi.org/10.1088/1475-7516/2017/11/012} {\bibfield  {journal}
  {\bibinfo  {journal} {JCAP}\ }\textbf {\bibinfo {volume} {1711}}\bibfield
  {number} {\bibinfo  {number} { (11)},\ \bibinfo {pages} {012}},\ }\bibinfo
  {note} {[JCAP1711,012(2017)]},\ \Eprint {https://arxiv.org/abs/1706.09895}
  {arXiv:1706.09895 [hep-ph]} \BibitemShut {NoStop}%
\bibitem [{\citenamefont {Connolly}\ \emph {et~al.}(2011)\citenamefont
  {Connolly}, \citenamefont {Thorne},\ and\ \citenamefont
  {Waters}}]{Connolly:2011vc}%
  \BibitemOpen
  \bibfield  {author} {\bibinfo {author} {\bibfnamefont {A.}~\bibnamefont
  {Connolly}}, \bibinfo {author} {\bibfnamefont {R.~S.}\ \bibnamefont
  {Thorne}},\ and\ \bibinfo {author} {\bibfnamefont {D.}~\bibnamefont
  {Waters}},\ }\bibfield  {title} {\bibinfo {title} {{Calculation of High
  Energy Neutrino-Nucleon Cross Sections and Uncertainties Using the MSTW
  Parton Distribution Functions and Implications for Future Experiments}},\
  }\href {https://doi.org/10.1103/PhysRevD.83.113009} {\bibfield  {journal}
  {\bibinfo  {journal} {Phys. Rev.}\ }\textbf {\bibinfo {volume} {D83}},\
  \bibinfo {pages} {113009} (\bibinfo {year} {2011})},\ \Eprint
  {https://arxiv.org/abs/1102.0691} {arXiv:1102.0691 [hep-ph]} \BibitemShut
  {NoStop}%
\bibitem [{\citenamefont {Arg$\ddot{\mathrm{u}}$elles}(2015)}]{delgado2015new}%
  \BibitemOpen
  \bibfield  {author} {\bibinfo {author} {\bibfnamefont {C.~A.}\ \bibnamefont
  {Arg$\ddot{\mathrm{u}}$elles}},\ }\emph {\bibinfo {title} {New physics with
  atmospheric neutrinos}},\ \href@noop {} {Ph.D. thesis},\ \bibinfo  {school}
  {The University of Wisconsin-Madison} (\bibinfo {year} {2015})\BibitemShut
  {NoStop}%
\bibitem [{\citenamefont {Jones}(2015)}]{jones2015sterile}%
  \BibitemOpen
  \bibfield  {author} {\bibinfo {author} {\bibfnamefont {B.~J.}\ \bibnamefont
  {Jones}},\ }\emph {\bibinfo {title} {{Sterile neutrinos in cold climates}}},\
  \href@noop {} {Ph.D. thesis},\ \bibinfo  {school} {MIT} (\bibinfo {year}
  {2015})\BibitemShut {NoStop}%
\bibitem [{\citenamefont {Collin}\ \emph {et~al.}(2016)\citenamefont {Collin},
  \citenamefont {Argüelles}, \citenamefont {Conrad},\ and\ \citenamefont
  {Shaevitz}}]{Collin:2016aqd}%
  \BibitemOpen
  \bibfield  {author} {\bibinfo {author} {\bibfnamefont {G.}~\bibnamefont
  {Collin}}, \bibinfo {author} {\bibfnamefont {C.}~\bibnamefont {Argüelles}},
  \bibinfo {author} {\bibfnamefont {J.}~\bibnamefont {Conrad}},\ and\ \bibinfo
  {author} {\bibfnamefont {M.}~\bibnamefont {Shaevitz}},\ }\bibfield  {title}
  {\bibinfo {title} {{First Constraints on the Complete Neutrino Mixing Matrix
  with a Sterile Neutrino}},\ }\href
  {https://doi.org/10.1103/PhysRevLett.117.221801} {\bibfield  {journal}
  {\bibinfo  {journal} {Phys. Rev. Lett.}\ }\textbf {\bibinfo {volume} {117}},\
  \bibinfo {pages} {221801} (\bibinfo {year} {2016})},\ \Eprint
  {https://arxiv.org/abs/1607.00011} {arXiv:1607.00011 [hep-ph]} \BibitemShut
  {NoStop}%
\bibitem [{\citenamefont {Jeffreys}(1998)}]{jeffreys1998theory}%
  \BibitemOpen
  \bibfield  {author} {\bibinfo {author} {\bibfnamefont {H.}~\bibnamefont
  {Jeffreys}},\ }\href@noop {} {\emph {\bibinfo {title} {The theory of
  probability}}}\ (\bibinfo  {publisher} {OUP Oxford},\ \bibinfo {year}
  {1998})\BibitemShut {NoStop}%
\end{thebibliography}%
\end{document}